\documentclass[12pt,epsfig]{article}
\usepackage[dvips]{graphicx}
\usepackage{amsfonts}
\usepackage{amsmath}
\usepackage{amssymb}

\newtheorem{postulate}{Postulate}[section]
\newcommand{\bpost}{\begin{postulate} \rm}
\newcommand{\epost}{\end{postulate}}
\newtheorem{theorem}{Theorem}[section]
\newcommand{\btheo}{\begin{theorem} \it}
\newcommand{\etheo}{\end{theorem}}
\newtheorem{proposition}{Proposition}[section]
\newcommand{\bprop}{\begin{proposition}\it}
\newcommand{\eprop}{\end{proposition}}
\newtheorem{corollary}{Corollary}[section]
\newcommand{\bcorol}{\begin{corollary} \it}
\newcommand{\ecorol}{\end{corollary}}
\newtheorem{lemma}{Lemma}[section]
\newcommand{\blem}{\begin{lemma}\it}
\newcommand{\elem}{\end{lemma}}
\newtheorem{remark}{Remark}[section]
\def\brem{\begin{remark} \rm }
\def\erem{\end{remark}\ \\}
\newtheorem{example}{Example}[section]
\def\bexam{\begin{example} ~~\\ \rm}
\def\eexam{\end{example}}

\newcommand{\bdefi}{\begin{definition} \rm }
\newcommand{\edefi}{\end{definition}\ \\}

\newcommand{\half}{\frac{1}{2}}

\usepackage{color}

\begin{document}

\date{}

\title{\textbf{Condensation in
a Disordered Infinite-Range Hopping Bose-Hubbard Model}}
\maketitle

\begin{center}

\setcounter{footnote}{0}
\renewcommand{\thefootnote}{\arabic{footnote}}

\vspace{1cm} {\bf T.C.DORLAS\footnote{DIAS-School of Theoretical
Physics - email: dorlas@stp.dias.ie},
L.A.PASTUR\footnote{Department of Theoretical Physics of ILTPE -
email: pastur@ilt.kharkov.ua} and V.A.
ZAGREBNOV\footnote{Universit\'e de la M\'editerran\'ee
(Aix-Marseille II) - email : zagrebnov@cpt.univ-mrs.fr}}

\vspace{0.5cm}

{\bf Dublin Institute for Advanced Studies, School of Theoretical
Physics,\\
10 Burlington Road, Dublin 4, Ireland$^{1}$}

\vspace{0.3cm}

{\textbf{Department of Theoretical Physics of Institute for Low
Temperature Physics and Engineering\\
47 Lenin Avenue, Kharkov 310164,  Ukraine $^{2}$}}

\vspace{0.3cm}

{\bf Universit\'e de la M\'editerran\'ee and Centre de Physique
Th\'eorique, \\Luminy-Case 907, 13288 Marseille, Cedex 09,
France$^{3}$}

\vspace{1.5cm}
\end{center}

\begin{abstract}
We study  Bose-Einstein Condensation (BEC) in the Infinite-Range
Hopping Bose-Hubbard  model for repulsive on-site particle
interaction in presence of ergodic random one-site potentials with
different distributions. We show that the model is exactly soluble
even if the on-site interaction is random. But in contrast to the
non-random case \cite{BD}, we observe here new phenomena: instead
of \textit{enhancement} of BEC for perfect bosons, for constant
on-site repulsion and discrete distributions of the single-site
potential there is \textit{suppression} of BEC at some
\textit{fractional} densities. We show that this suppression
appears with increasing disorder. On the other hand, the BEC
suppression at integer densities may disappear, if disorder
increases. For a continuous distribution we prove that the BEC
critical temperature decreases for small on-site repulsion while
the BEC is suppressed at integer values of density for large
repulsion. Again, the threshold for this repulsion gets higher,
when disorder increases.
\end{abstract}

\newpage

\section{Introduction}
\setcounter{equation}{0} \renewcommand{\theequation}{\arabic{section}.\arabic{equation}}
\setcounter{theorem}{0}


Lattice Bose-gas models were invented as an alternative way to
understand  continuous interacting boson systems including liquid
Helium, see \cite{Mat-Mat} and a very complete review
\cite{Ueltschi}. But recent experiments with cold bosons in traps of
three-dimensional optical lattice potentials show that lattice
models are also relevant for describing the experimentally observed
Mott \textit{insulator}-\textit{superfluid} (or condensate) phase
transition \cite{GMEHB}. In \cite{BD} and then in \cite{ALSSY}, this
phenomenon was analyzed rigorously in the framework of the so-called
\textit{Bose-Hubbard} model.

The aim of the present paper is to study a \textit{disordered}
Bose-Hubbard model and in particular the influence of the
\textit{single-site potential} randomness on the Bose-Einstein
condensate (BEC).

Notice that the first attempts to understand this influence go back
to \cite{KL1}, \cite{KL2} and \cite{LS} for continuous Perfect
Bose-Gases (PBG) in a random potential of \textit{impurities}. For
the rigorous solution of this problem see \cite{LPZ}. One of the
principal result of \cite{LPZ} is that the randomness
\textit{enhances} the BEC. For example, the one-dimensional PBG has
no BEC because of the high value of the one-particle density of
states in the vicinity of the bottom of the spectrum above the
ground state, making the integral for the critical particle density
infinite. The presence of a non-negative homogeneous ergodic random
potential modifies the one-particle density of states (due to the
\textit{Lifshitz tail}) in such a way that the integral for the
critical density becomes finite. Hence, the one-dimensional PBG with
random potential does manifest BEC. The nature of this BEC is close
to what is known as the "\textit{Bose-glass}" since it may be
localized by the random potential \cite{LZ}. This is of interest for
experiments with liquid $^4He$ in random environments like Aerogel
and Vycor glass, \cite{FWGF}, \cite{KT}.

On the other hand, the nature and behaviour of the \textit{lattice}
BEC may be quite different. First of all, the lattice Laplacian and
the Bose-Hubbard interaction produce a coexistence of the BEC
(\textit{superfluidity}) and the \textit{Mott insulating phase} as
well as domains of \textit{incompressibility}, see e.g. \cite{FWGF},
\cite{KTC}. Adding disorder makes the corresponding models much more
complicated. The physical arguments \cite{FWGF}, \cite{KTC} show
that the randomness may  \textit{suppress} the BEC (superfluidity)
as well as the Mott phase in favour of the localized
\textit{Bose-glass} phase, but this is very sensitive to the choice
of the random distribution.

Since there are very few rigorous results about the BEC in
disordered systems, we consider here a \textit{single-site} random
version of the lattice  \textit{Infinite-Range Hopping} (IRH)
Bose-Hubbard model, which in non-random  case has recently been
studied in detail for all temperatures and chemical potentials in
\cite{BD}.

This paper is organized as follows. In Section~2, we define the
lattice Laplacian for finite- and infinite-range hopping and recall
the results about BEC for the free lattice Bose-gas. We then
introduce random single-site and on-set particle interaction
potentials and state our main result about the existence of and an
explicit formula for the pressure for the IRH Bose-Hubbard model
with these type of randomness. We outline the proof of the main
theorem using the approximating Hamiltonian method.

In Section~3 we consider the pressure for extremal cases of
\textit{hard-core} and \textit{perfect} bosons. We show that they
are the limits of the IRH Bose-Hubbard model pressure when the
on-site particle interaction tends respectively to $+\infty$ and to
$0$.

In Section 4, we analyse the phase diagram in the case of a
non-random on-site particle interaction and random single-site
external potential. We distinguish a number of different cases. We
start with \textit{perfect} bosons and show that the randomness
\textit{enhances} BEC in this case, see Sect.4.1. This is no longer
true for  interacting bosons. We study in Sect.4.2 the phase diagram
first for \textit{Bernoulli} single-site potential and then for
\textit{trinomial} and \textit{multinomial} discrete distributions.

In the case of a Bernoulli distribution and hard-core bosons
(infinite on-set repulsion) we showe that in addition to the
complete BEC \textit{suppression} at extremal allowed densities
$\rho=0$ and $\rho=1$ there is a new point $\rho=1 -p$, where $p = \
$Pr$\,\{potential \neq 0 \}$. We prove that for  finite on-site
repulsion the suppression of BEC  at \textit{integer}, and also for
\textit{fractional} values of densities $\rho=n-p \ , n = 1, 2,
\ldots $ persists, if the Bernoulli potential amplitude is large
enough. In fact we find that increasing the Bernoulli potential
amplitude (disorder) decreases the critical BEC temperature in the
vicinity of fractional values of densities but increases it for
integer values of density. A similar phenomenon occurs also for
\textit{equiprobable} trinomial distributions, but now for densities
$\rho=n/3$. Our numerical calculations demonstrate that it should be
true for a general multinomial distribution.

For illustration of a continuous distribution we study a homogenous
distribution with  compact support. Then for hard-core bosons we
prove that the complete BEC suppression occurs \textit{only} at
extremal allowed densities $\rho=0$ and $\rho=1$, with the trace of
suppressions only at \textit{integer} values of densities for a
finite on-site repulsion. In particular we show that the critical
BEC temperature gets \textit{lower}, when one switches on disorder
for (a small) on-site interaction, whereas it gets \textit{higher}
for perfect bosons. For large values of on-site interaction the
picture is similar to the discrete distributions: increasing of
disorder increases the critical BEC temperature in the vicinity of
integer values of density but increases it for the complimentary
values of density.

In Section 5 we summarize and discuss our results.

\section{Model and Main Theorem}

\setcounter{equation}{0} \renewcommand{\theequation}{\arabic{section}.\arabic{equation}}

For simplicity we shall consider the Bose-Hubbard model only with
\textit{periodic boundary conditions}. So let $\Lambda:= \{x\in
{\mathbb{Z}}^{d}: -L_\alpha/2 \leq x_\alpha < L_\alpha/2  , \
\alpha=1,\ldots,d \}$ be a bounded rectangular domain of the cubic
lattice ${\mathbb{Z}}^{d}$ wrapped onto a \textit{torus}. Then the
set $\Lambda^* := \{q_\alpha = {2\pi n}/{L_\alpha}: n =
0,\pm1,\pm2, \ldots \pm(L_\alpha/2-1), L_\alpha/2, \ \alpha = 1,
2,\ldots d \}$ is  \textit{dual} to $\Lambda$ with respect to
Fourier transformation on the domain $\Lambda = L_1 \times L_2
\times \ldots \times L_d$ of volume  $\left|\Lambda\right| = V$.

The standard \textit{one-particle} Hilbert space for the set $\Lambda$ can be taken as
$\mathfrak{h}(\Lambda):= \mathbb{C}^\Lambda$ with the canonical
basis $\{e_x \}_{x\in\Lambda}$, i.e. $e_x(y)= \delta_{x\,,\,y}$.
Then for any element $u = \sum_{x\in \Lambda}u_x e_x \in
\mathfrak{h}(\Lambda)$ the one-particle \textit{kinetic-energy}
(\textit{hopping}) operator is defined by
\begin{equation}\label{one-kinetic-energy}
(t_{\Lambda}u)(x):= \sum_{y \in \Lambda} t_{x\,,\,y}^{\Lambda} (u(x) - u(y))=
\sum_{y \in \Lambda} t_{x\,,\,y}^{\Lambda} (u_x - u_y),
\end{equation}
where
\begin{equation}\label{hopping-matr}
t_{x\,y}^{\Lambda} = \frac{1}{V}\sum_{q\in\Lambda^*} \hat{t}_q e^{i q(x-y)} \ , \
\end{equation}
is the \textit{periodic extension} in domain $\Lambda$ of a \textit{symmetric}, \textit{translation
invariant} and \textit{positive-definite} matrix, i.e.
\begin{equation}\label{hopping-matr}
\hat{t}_q =  \sum_{y\in\Lambda}t_{0\,,\,y}^{\Lambda} e^{i q y} \geq 0 .
\end{equation}
Notice that functions $\{ (\hat{e}_q)(y):= e^{iqy}/\sqrt{V}\}_{q\in\Lambda^*}$ also form a basis
in $\mathfrak{h}(\Lambda)$, i.e. for any $u \in \mathfrak{h}(\Lambda)$ one has
$u = \sum_{q\in \Lambda^*}u_q \hat{e}_q $ .

Let $\mathfrak{F}_B:=\mathfrak{F}_B(\mathfrak{h}(\Lambda))$ be the \textit{boson Fock space} over
$\mathfrak{h}(\Lambda)$. For any $f\in \mathfrak{h}(\Lambda))$ we can associate in this space
the creation  and annihilation operators
\begin{equation}\label{creat-annih-oper}
a^{*}(f):=\sum_{y\in\Lambda} a^{*}(y) f(y) \ , \  \ a(f):=\sum_{y\in\Lambda} a(y) f^{*}(y) \ .
\end{equation}
Let  $a_{x}^{*}$, $a_{x}$ and $\hat{a}_{q}^{*}$, $\hat{a}_{q}$ be
the boson creation and annihilation operators corresponding respectively to the basis elements
$e_x$ and $\hat{e}_q$, satisfying the lattice \textit{Canonical Commutation
Relations}: $\left[a_{x},a_{y}^{*}\right]=\delta_{x\,,\,y}$  and
$\left[\hat{a}_{q},\hat{a}_{p}^{*}\right]=\delta_{q\,,\,p}$. Then $n_{x}=a_{x}^{*}a_{x}$ is
the \textit{one-site} number operator, and
\begin{equation}\label{N-oper}
N_{\Lambda}:= \sum_{x \in\Lambda}n_{x} = \sum_{q \in\Lambda^*}
\hat{a}_{q}^*\hat{a}_q\,,
\end{equation}
is the \textit{total} number operator.

The second quantization of the hopping operator (\ref{one-kinetic-energy}) in $\mathfrak{F}_B$
gives  the \textit{free boson} Hamilton of the form
\begin{equation}\label{free-Ham}
T_{\Lambda}:= \sum_{x\, \in \Lambda} a^{*}_{x} (t_{\Lambda}a)_x =
\frac{1}{2} \sum_{x,y \, \in \Lambda} t_{x\,y}^{\Lambda} (a^{*}_{x}-a^{*}_{y})(a_{x}-a_{y})
= \sum_{q \in \Lambda^*} (\hat{t}_{0}-\hat{t}_{q}) \hat{a}^{*}_{q} \hat{a}_{q}.
\end{equation}
If hopping is allowed only between the \textit{nearest neighbor}
(n.n.) sites with equal probabilities, then $t_{\Lambda} = -
\Delta $ corresponds to minus the \textit{lattice Laplacian}, i.e.
\begin{equation}\label{hopp-n-n} t_{x\,y}^{\Lambda} = \sum_{\alpha
= 1}^d \ (\delta_{x+ 1_\alpha \,,\, y} +  \delta_{x - 1_\alpha
\,,\, y}),
\end{equation}
where $(x \pm 1_\alpha)_\beta = x_\beta \pm \delta_{\alpha
\,,\,\beta}$. In this case the one-particle hopping operator
spectrum is
\begin{equation}\label{spect-n-n}
\epsilon(q):=(\hat{t}_{0}-\hat{t}_{q}) = \sum_{\alpha = 1}^d \ 4 \
\sin^2(q_\alpha/2) \geq 0 \ , \ \ q \in \Lambda^* \ ,
\end{equation}
with eigenfunctions $\{\hat{e}_q\}_{q \in \Lambda^*}$.

It is known that the lattice free Bose-gas (\ref{free-Ham}) with
\textit{n.n.} hopping manifests the \textit{zero-mode} BEC when $d >
2$, since the spectral density of states
$\mathcal{N}_{d}(d\epsilon)$ corresponding to (\ref{hopp-n-n}) is
small enough to make the \textit{critical} particle density
$\rho_{c}^{free}(\beta)$ \textit{bounded} for a given temperature
$\beta ^{-1}$ :
\begin{eqnarray}\label{cr-dens-free-nn}
\rho_{c, \ n.n.}^{free}(\beta):= \lim_{\mu \uparrow 0}
\lim_{\Lambda} \frac{1}{V} \sum_{q \in \Lambda^*} \
\frac{1}{e^{\beta(\epsilon(q)-\mu)}-1} &=&
\frac{1}{(2\pi)^d}\int_{\mathcal{B}^d} d^{d}q \ \frac{1}{e^{\beta
\epsilon(q)}-1}\\
\nonumber &=& \int_{\mathbb{R}_{+}} \ \mathcal{N}_{d}(d\epsilon) \
\frac{1}{e^{\beta \epsilon}-1} < \infty \ .
\end{eqnarray}
Here $\lim_{\Lambda}$ stands for the thermodynamic limit $\Lambda
\uparrow \mathbb{Z}^d$, by $\mathcal{B}^d := [-\pi,\pi]^d$ we denote the first
\textit{Brillouin zone} and the density of states
$\mathcal{N}_{d}(d\epsilon) = \{c_d \epsilon^{(d/2-1)}+
{o}(\epsilon^{(d/2-1)})\}d\epsilon$ for small $\epsilon$.

A similar result is true for the \textit{infinite-range} (i.r.) hopping Laplacian:
\begin{equation}\label{hopp-inf-rang}
t_{x\,y}^{\Lambda} = \frac{1}{V} (1 - \delta_{x\,,\,y}) \ , \ x,y
\in \Lambda.
\end{equation}
 By (\ref{hopp-inf-rang}) the one-particle spectrum in this case takes the form:
\begin{equation}\label{spect-inf-rag}
\epsilon(q):=(\hat{t}_{0}-\hat{t}_{q}) = (1 - \delta_{q\,,\,0}) \geq
0 \ , \ \ q \in \Lambda^* \ .
\end{equation}
Therefore, it has a \textit{gap}:
\begin{equation}\label{gap}
\lim_{q\rightarrow 0} \epsilon(q)= 1 \neq \epsilon(0) = 0 \ ,
\end{equation}
and allowed values of the chemical potential are still $\mu \leq 0$.
Since the density of states is simply zero in the gap, and $|\Lambda^*| = V  |\mathcal{B}^d|$, we have
$\mathcal{N}_{d}(d\epsilon) =  \delta (\epsilon - 1)
d\epsilon$. Therefore, the \textit{critical} particle density has
a bounded  value:
\begin{equation}\label{cr-dens-free-ir}
\rho_{c, \ i.r.}^{free}(\beta)=
\frac{1}{(2\pi)^d}\int_{\mathcal{B}^d} d^{d}q \ \frac{1}{e^{\beta}-1}
=  \
\frac{1}{e^{\beta}-1} < \infty \ ,
\end{equation}
for any dimensions. The latter implies a zero-mode BEC for densities $\rho > \rho_{c, \ i.r.}^{free}(\beta)$.

The problem of existence of BEC gets much less obvious if one
takes into account the \textit{boson interaction}. This is even
the case for the simplest \textit{on-site} repulsive interaction
\begin{equation}\label{Bose-Hubb}
H_{\Lambda}:= T_{\Lambda} + \lambda \ \sum_{x \in\Lambda} n_{x} (n_{x}-1) \ , \ \lambda \geq 0 \ ,
\end{equation}
known as the \textit{Bose-Hubbard} model. (Notice that \textit{attraction}:  $\lambda
<0$ makes this model unstable, see \cite{Ueltschi} for discussion
of other cases.)
\begin{remark}\label{best-results}
Concerning the model (\ref{Bose-Hubb}) the best rigorous results so far are: \\
- a proof of BEC for the n.n. lattice Laplacian and the
\textit{hard-core} boson repulsion: $\lambda = +\infty$, by
{\rm{\cite{KLS}}} for the case of the half-filled lattice, see also {\rm{\cite{AB}}}; \\
- a recent exact solution of the IRH Bose-Hubbard model (\ref{hopp-inf-rang}),
(\ref{Bose-Hubb}) for any $\lambda \geq 0$ by {\rm{\cite{BD}}}.
\end{remark}

The aim of the the present paper is to study a \textit{disordered} IRH Bose-Hubbard model.
Let $\left(\Omega,\Sigma,\mathbb{P} \right)$ be a probability
space. We define our basic model by the random Hamiltonian:
\begin{equation}\label{ham-1}
H_{\Lambda}^{\omega}= \frac{1}{2 V}\sum_{x,y
\in\Lambda}(a^{*}_{x}-a^{*}_{y})(a_{x}-a_{y}) + \sum_{x
\in\Lambda}\lambda_{x}^\omega n_{x}(n_{x}-1) + \sum_{x
\in\Lambda}\varepsilon_{x}^{\omega}n_{x},
\end{equation}
where parameters
$\left\{\lambda_{x}^{\omega} \geq 0\right\}_{{x}\in{\mathbb{Z}}^{d}}$ and
$\left\{\varepsilon_{x}^{\omega}\in \mathbb{R}^1\right\}_{{x}\in{\mathbb{Z}}^{d}}$,
for $\omega\in\Omega$, are real-valued random fields on
${\mathbb{Z}}^{d}$, which we suppose to be \textit{stationary} and
\textit{ergodic}. We denote by
\begin{equation}\label{press}
p_{\Lambda}^{\omega}(\beta,\mu):= p\left[H_{\Lambda}^{\omega}
\right](\beta,\mu):= \frac{1}{\beta V} \mbox{Tr}_{\mathfrak{F}_B}
\exp\left\{-\beta(H_{\Lambda}^{\omega}-\mu N_\Lambda)\right\}
\end{equation}
the grand canonical pressure of the system (\ref{ham-1}) for given
temperature $\beta^{-1}$ and chemical potential $\mu$. For \textit{non-random}
parameters $\lambda_{x}^\omega = \lambda \geq 0$ and
$\varepsilon_{x}^{\omega}= \varepsilon =0$ the model (\ref{ham-1}) was considered in \cite{BD}.

Our main theorem is a formula for the pressure of this model given
some general regularity conditions on the random parameters
involved in the Hamiltonian (\ref{ham-1}).
\begin{theorem}\label{main-1}
Let the stationary, ergodic random fields
$\left\{\lambda_{x}^{\omega}\right\}_{x\in{\mathbb{Z}}^{d}}$ and
$\left\{\varepsilon_{x}^{\omega}\right\}_{x\in{\mathbb{Z}}^{d}}$
be such that:
\begin{equation}\label{hyp-on-rand}
\lambda_{\rm min}:=\inf_{x, \omega}\lambda_{x}^{\omega} > 0 \,\,\,\,,\, \ \ \
\,\,\,\,\,\,  \varepsilon_{\rm min}:=\inf_{x, \omega}\varepsilon_{x}^{\omega} \ > -\infty.
\end{equation}
Then for almost all $\omega\in\Omega$, i.e., almost sure {\rm{(a.s.)}}, there exists a non-random
thermodynamic limit of the pressure (\ref{press}):
\begin{equation}\label{lim-press-1}
a.s.-\lim_{\Lambda}p_{\Lambda}^{\omega}(\beta,\mu)=p(\beta,\mu),
\end{equation}
such that
\begin{eqnarray}\label{lim-press-2}
&& p(\beta,\mu)=  \\
\nonumber &&\sup_{r\geq0}\left\{-r^{2}+
\beta^{-1}\mathbb{E}\left\{\ln \mbox{{\rm{Tr}}}_{({\mathfrak{F}_B})_x}
\exp\beta\left[(\mu-\varepsilon_{x}^{\omega}-1)n_{x}-
\lambda_{x}^\omega n_{x}(n_{x}-1)+ r (a^{*}_{x} +
a_{x})\right]\right\}\right\} ,
\end{eqnarray}
where $\mathbb{E}\left(\cdot\right)$ is expectation with respect to
the measure $\mathbb{P}$.
\end{theorem}
\noindent \textit{Proof}: Let
\begin{equation}\label{appr-ham}
H_{0\Lambda}^{\omega}:= \sum_{x \in\Lambda}\lambda_{x}^\omega
n_{x}(n_{x}-1) + \sum_{x
\in\Lambda}(\varepsilon_{x}^{\omega}+1)n_{x} \ .
\end{equation}
Then by definitions (\ref{creat-annih-oper}) the Hamiltonian (\ref{ham-1})
takes the form
\begin{equation}\label{ham-2}
H_{\Lambda}^{\omega}= T_{\Lambda} + \sum_{x
\in\Lambda}\lambda_{x}^\omega n_{x}(n_{x}-1) + \sum_{x
\in\Lambda}\varepsilon_{x}^{\omega}n_{x}= - \hat{a}_0 ^{*}\hat{a}_0
+ H_{0\Lambda}^{\omega} \ .
\end{equation}
Since conditions (\ref{hyp-on-rand}) imply the estimate from below:
\begin{eqnarray}\label{super-stab}
H_{\Lambda}^{\omega}&\geq& - \hat{a}_0 ^{*}\hat{a}_0 + N_{\Lambda}
+ \lambda_{\rm min} \ \sum_{x \in\Lambda}
n_{x}(n_{x}-1)+ \varepsilon_{\rm min} N_{\Lambda} \\
\nonumber &\geq& \frac{\lambda_{\rm min}}{V}N_{\Lambda}^2 +
(\varepsilon_{\rm min} - \lambda_{\rm min})N_{\Lambda} \ ,
\end{eqnarray}
the Hamiltonian (\ref{ham-2}) is \textit{superstable}. Thus, the
pressure in (\ref{lim-press-1}) is defined for \textit{all} $\mu \in
\mathbb{R}^1$.

Following \cite{BBZKT}, we introduce a similar Hamiltonian with
\textit{sources}:
\begin{equation}\label{ham-nu}
H_{\Lambda}^{\omega}(\nu):= H_{\Lambda}^{\omega} -
\sqrt{V}(\overline{\nu}\hat{a}_0 + \nu \hat{a}_0 ^{*})\,,\,\,\,
\nu \in \mathbb{C} \,,
\end{equation}
and the corresponding \textit{approximating} Hamiltonian:
\begin{equation}\label{ham-appr-nu}
H_{\Lambda}^{\omega}(z,\nu):= H_{0 \Lambda}^{\omega}(z) -
\sqrt{V}( \overline{\nu} \hat{a}_0 + \nu \hat{a}_0 ^{*} )\,,
\end{equation}
where
\begin{equation}\label{ham-appr}
H_{0 \Lambda}^{\omega}(z):= H_{0\Lambda}^{\omega} +
V\left|z\right|^2 - \sqrt{V}( \overline{z} \hat{a}_0 + z \hat{a}_0
^{*} ) \,\,,\,\,\, z \in \mathbb{C}.
\end{equation}
Then
\begin{equation}\label{diff-ham-ham-appr}
H_{\Lambda}^{\omega}(\nu)- H_{\Lambda}^{\omega}(z,\nu)=
-(\hat{a}_0 - z\sqrt{V})^*(\hat{a}_0 - z\sqrt{V}),
\end{equation}
and by virtue of the Bogoliubov \textit{convexity inequality} one
gets the estimates:
\begin{equation}\label{Bog-ineq-1}
 0 \leq p\left[H_{\Lambda}^{\omega}(\nu)\right] - p\left[H_{\Lambda}^{\omega}(z,\nu)
\right]\leq \frac{1}{V}\left\langle(\hat{a}_0 -
z\sqrt{V})^*(\hat{a}_0 - z\sqrt{V})
\right\rangle_{H_{\Lambda}^{\omega}(\nu)}
\end{equation}
for \textit{each} realization $\omega\in\Omega$. Here
$\left\langle - \right\rangle_{H_{\Lambda}^{\omega}(\nu)}:=
\left\langle - \right\rangle_{H_{\Lambda}^{\omega}(\nu)} (\beta,
\mu)$ denotes the grand-canonical quantum Gibbs state with
Hamiltonian (\ref{ham-nu}), and from now on we systematically omit
the arguments $(\beta,\mu)$. If we choose in the right-hand side
of (\ref{Bog-ineq-1})
\begin{equation}\label{put-z}
z = \frac{1}{\sqrt{V}}\left\langle
\hat{a}_0\right\rangle_{H_{\Lambda}^{\omega}(\nu)},
\end{equation}
then (\ref{Bog-ineq-1}) implies the following estimate for each
$\omega\in\Omega$:
\begin{equation}\label{Bog-ineq-2}
 0 \leq p\left[H_{\Lambda}^{\omega}(\nu)\right] -
 \sup_{z \in \mathbb{C}} p\left[H_{\Lambda}^{\omega}(z,\nu)
\right]\leq \frac{1}{V}\left\langle \delta\hat{a}_0^*\,
\delta\hat{a}_0 \right\rangle_{H_{\Lambda}^{\omega}(\nu)},
\end{equation}
where we denote
\begin{equation}\label{dev-a}
\delta\hat{a}_0 := \hat{a}_0 - \left\langle
\hat{a}_0\right\rangle_{H_{\Lambda}^{\omega}(\nu)}.
\end{equation}
Since (\ref{N-oper}) implies the estimates:
\begin{equation}\label{estim-source}
- \sqrt{V}(\overline{\nu}\hat{a}_0 + \nu \hat{a}_0 ^{*})\geq -
\left|\nu \right|^{2} \hat{a}_0 ^{*}\hat{a}_0 - V  \geq - {\left|\nu
\right|^2} N_{\Lambda} - V,
\end{equation}
by virtue of (\ref{super-stab}) and  (\ref{estim-source}) the
Hamiltonian with sources (\ref{ham-nu}) is also
\textit{superstable}:
\begin{equation}\label{super-st}
H_{\Lambda}^{\omega}(\nu) \geq \frac{\lambda_{\rm
min}}{V}N_{\Lambda}^2  + (\varepsilon_{\rm min} -\lambda_{\rm min}
- \left|\nu \right|^2) \ N_{\Lambda} - V \ ,
\end{equation}
uniformly in  $\omega\in\Omega$ and in $\left|\nu \right|\leq C_0$,
for a fixed $C_0 \geq 0$. The superstability (\ref{super-st})
implies that there is a \textit{monotonous nondecreasing} function
$M := M(\beta, \mu)\geq 0$ of $\mu\in \mathbb{R}^1$, such that for
any $\omega\in\Omega$  we have the bounds:
\begin{eqnarray}
&&\left|\left\langle \hat{a}_0/{\sqrt{V}}
\right\rangle_{H_{\Lambda}^{\omega}(\nu)}(\beta, \mu)\right|^2 =
\left|\partial_{\overline{\nu}}
\,p\left[H_{\Lambda}^{\omega}(\nu)\right](\beta, \mu)\right|^2 \nonumber\\
 &&\leq \left\langle N_{\Lambda}/{V}
\right\rangle_{H_{\Lambda}^{\omega}(\nu)}(\beta, \mu)=
\partial_\mu \,p\left[H_{\Lambda}^{\omega}(\nu)\right](\beta,
\mu)\leq M^2, \label{bound-1}
\end{eqnarray}
and
\begin{equation}\label{bound-2}
\left|z_{\Lambda,\omega}(\beta, \mu; \nu)\right|^2\leq M^2
\end{equation}
for the \textit{maximizer} $z_{\Lambda,\omega}(\nu):=
z_{\Lambda,\omega}(\beta, \mu; \nu)$ in (\ref{Bog-ineq-2}):
\begin{equation}\label{max}
p\left[H_{\Lambda}^{\omega}(z_{\Lambda,\omega}(\beta, \mu;
\nu),\nu)\right](\beta, \mu):= \sup_{z\in\mathbb{C}}
p\left[H_{\Lambda}^{\omega}(z, \nu)\right](\beta, \mu) \ ,
\end{equation}
uniform in  $\left|\nu \right|\leq C_0$. Notice that the maximizer
satisfies the equation:
\begin{equation}\label{ident}
z_{\Lambda,\omega}(\nu) =
\partial_{\overline{\nu}}p\left[H_{\Lambda}^{\omega}(z_{\Lambda,\omega}(\nu),\nu)\right]
= \left\langle \hat{a}_0/{\sqrt{V}}
\right\rangle_{H_{\Lambda}^{\omega}(z_{\Lambda,\omega}(\nu),\ \nu)}.
\end{equation}
Moreover, by the same line of reasoning as in \cite{ZB}, Ch.4 (see
also \cite{BD}) one gets that for $\left|\nu \right|< C_0$ there are
some $u=u(M)>0$ and $w=w(M)>0$ such that
\begin{equation}\label{mean-scal-prod-esti}
\left\langle \delta\hat{a}_0^*\,
\delta\hat{a}_0 \right\rangle_{H_{\Lambda}^{\omega}(\nu)}\leq
\left\{u + w (\delta\hat{a}_0^*\,,
\delta\hat{a}_0)_{H_{\Lambda}^{\omega}(\nu)}\right\},
\end{equation}
where
\begin{equation}\label{Bog-scal}
(\delta\hat{a}_0^*\,, \delta\hat{a}_0)_{H_{\Lambda}^{\omega}(\nu)}
= \beta^{-1}
\partial_{\nu}\,\partial_{\overline{\nu}}\,p\left[H_{\Lambda}^{\omega}(\nu)\right].
\end{equation}
Then the estimates (\ref{Bog-ineq-2}) and (\ref{mean-scal-prod-esti}) imply:
\begin{equation}\label{Bog-ineq-3}
0 \leq p\left[H_{\Lambda}^{\omega}(\nu)\right] -
p\left[H_{\Lambda}^{\omega}(z_{\Lambda,\omega}(\nu), \nu)
\right]\leq \frac{1}{V}\left\{u + w (\delta\hat{a}_0^*\,,
\delta\hat{a}_0)_{H_{\Lambda}^{\omega}(\nu)}\right\}.
\end{equation}

Following  \cite{PS} we define in the Hilbert space
$L^2(\left\{(\mbox{Re} \nu,\, \mbox{Im} \nu) \in \mathbb{R}^2
:\left|\nu\right|< C_0\right\})$ the \textit{Dirichlet}
self-adjoint extension $\hat{L}_V$  of the operator
\begin{equation}\label{oper}
{L}_{V} := I - w (\beta V)^{-1} \,
\partial_{\nu}\,\partial_{\overline{\nu}}\,\,.
\end{equation}
Here $4 \partial_{\nu}\,\partial_{\overline{\nu}} = \Delta$ coincides
with the two-dimensional Laplacian operator in variables $(\mbox{Re} \nu,\, \mbox{Im}
\nu)$. The operator $\hat{L}_V$ is invertible and $\hat{L}_{V}^{-1}$
has the kernel $\left(\hat{L}_{V}^{-1}\right)(\nu, \nu')$
(\textit{Green} function), and $\left(\hat{L}_{V}^{-1}\right)(\nu,
\nu')=0$ for $\left|\nu\right|= C_0$, or $\left|\nu'\right|= C_0$,
by the Dirichlet boundary condition. Since the semigroup $\left\{
\exp \ [-t(\hat{L}_{V}- I)]\right\}_{t\geq0}$ is
\textit{positivity preserving}, the same property is true for the operator
$\hat{L}_{V}^{-1}$, see e.g. \cite{RS II}, Ch.X.4.

Now, let $p (\nu) := p\left[H_{\Lambda}^{\omega}(\nu)\right]$ and
$p_0 (\nu):=
p\left[H_{\Lambda}^{\omega}(z_{\Lambda,\omega}(\nu),\nu)\right]$.
Since $\hat{L}_{V}^{-1}$ is positivity preserving, then
(\ref{Bog-ineq-3})-(\ref{oper}) imply
\begin{equation}\label{estim-1}
\left(\hat{L}_{V}^{-1}(p_0 + u/V)\right)(\nu) \geq p(\nu) \,,
\end{equation}
and by consequence the estimates
\begin{eqnarray}
&&0 \leq p\left[H_{\Lambda}^{\omega}(\nu)\right] -
p\left[H_{\Lambda}^{\omega}(z_{\Lambda,\omega}(\nu), \nu)
\right]\leq \left(\hat{L}_{V}^{-1}(p_0 + u/V )\right)(\nu) - p_0(\nu) \nonumber \\
&&\leq \int_{\left|\nu'\right|<
C_0}d\nu'\left(\hat{L}_{V}^{-1}\right)(\nu, \nu')\left\{p_0 (\nu') -
p_0(\nu)\right\} + u/V , \label{estim-2-0}
\end{eqnarray}
where we used that $\int_{\left|\nu'\right|<
C_0}d\nu'\left(\hat{L}_{V}^{-1}\right)(\nu, \nu')=1$,
$\left|\nu\right|< C_0$. By virtue of (\ref{bound-2}) and
(\ref{ident}) we obtain for the integral in the right-hand side of
(\ref{estim-2}) the estimate:
\begin{equation}\label{estim-3}
\int_{\left|\nu'\right| <
C_0}d\nu'\left(\hat{L}_{V}^{-1}\right)(\nu, \nu')\left\{p_0 (\nu') -
p_0(\nu)\right\}\leq 2 M \int_{\left|\nu'\right| <
C_0}d\nu'\left(\hat{L}_{V}^{-1}\right)(\nu, \nu')\left|\nu' -
\nu\right|= I_{V}.
\end{equation}
After change of variables to  $\xi = \nu \sqrt{V}$, we get
\begin{equation}\label{estim-4}
I_{V} =  \frac{2 M}{V}\int_{\left|\xi'\right| < C_0 \sqrt{
V}}d\xi'\left(\hat{L}_{V=1}^{-1}\right)(\xi, \xi')\left|\xi' -
\xi\right|\leq \frac{\tilde{M}}{V}.
\end{equation}
Here we used that in $\mathbb{R}^2$ the \textit{Green} function is
known explicitly:
\begin{equation}\label{Green-Bessel}
\left(\hat{L}_{\infty}^{-1}\right)(\xi, \xi') = \frac{w}{2 \pi
\beta} K_{0}(\frac{\beta}{w}\left|\xi - \xi' \right|) ,
\end{equation}
where the Bessel function $K_{0}(x) \simeq \sqrt{\pi/2x} \exp(- x)$
decays exponentially fast for large $x > 0$. Therefore,
(\ref{estim-2-0}) and (\ref{estim-4}) imply
\begin{equation}\label{lim-1}
0 \leq p\left[H_{\Lambda}^{\omega}(\nu)\right] -
p\left[H_{\Lambda}^{\omega}(z_{\Lambda,\omega}(\nu), \nu)
\right]\leq O(1/V) \ ,
\end{equation}
for all $\omega\in\Omega$, any $\beta>0$, $\mu\in \mathbb{R}^1$ and $\left|\nu\right|< C_0$.

Notice that by definitions (\ref{appr-ham})
and (\ref{ham-appr}) for any $z, \nu \in \mathbb{C}$ we get:
\begin{eqnarray}\label{appr-press}
&& p_{\Lambda,\,\rm{appr}}^{\omega}(\beta, \mu; z, \nu):=
p\left[H_{\Lambda}^{\omega}(z,\nu)\right](\beta, \mu) =
-\left|z\right|^{2} +
\\
&& + \frac{1}{\beta V}\sum_{x\in\Lambda}\ln {\mbox{Tr}}_{{\mathfrak{F}}_x}
\exp \beta \left[(\mu-\varepsilon_{x}^{\omega}-1)n_{x}-
\lambda_{x}^{\omega} n_{x}(n_{x}-1)+  (z+\nu){a}^{*}_{x} +
(\overline{z}+\overline{\nu}) {a}_{x}\right] .\nonumber
\end{eqnarray}
Then ergodicity of the random fields
$\left\{\lambda_{x}^{\omega}\right\}_{x\in{\mathbb{Z}}^{d}}$ and
$\left\{\varepsilon_{x}^{\omega}\right\}_{x\in{\mathbb{Z}}^{d}}$
implies the existence of the \textit{a.s.} limit:
\begin{eqnarray}\label{lim-appr-press}
&& p_{\,\,\rm{appr}}(\beta, \mu; z, \nu) = a.s.-\lim_{\Lambda}
p_{\Lambda,\,\,\rm{appr}}^{\omega}(\beta, \mu; z, \nu)= -
\left|z\right|^{2} +
\\
&& + \beta^{-1}\mathbb{E}\left\{\ln {{\mbox{Tr}}}_{{\mathfrak{F}}_x}
\exp \beta \left[(\mu-\varepsilon_{x}^{\omega}-1)n_{x}-
\lambda_{x}^{\omega} n_{x}(n_{x}-1)+ (z+\nu){a}^{*}_{x} +
(\overline{z}+\overline{\nu}) {a}_{x}\right] \right\}, \nonumber
\end{eqnarray}
i.e., the \textit{self-averaging} \cite{PF} of the limiting
approximating pressure $p_{\,\,\rm{appr}}^{\omega}(\beta, \mu; z,
\nu)$.

Now we put the source $\nu \rightarrow 0$ and we make the canonical
(\textit{gauge}) transformation:
\begin{equation}\label{gauge}
\tilde{a}_x := a_x e^{i\,arg\,z}.
\end{equation}
Since Hamiltonian (\ref{ham-appr}) is invariant with respect of
this transformation, we get that $z = \left|z\right|:= r$ and
(cf.(\ref{appr-press})):
\begin{eqnarray}\label{appr-press-1}
&& \tilde{p}_{\Lambda,\,\rm{appr}}^{\omega}(\beta, \mu;
r):=p_{\Lambda,\,\rm{appr}}^{\omega}(\beta, \mu; z=r, \nu =0)=
p\left[H_{\Lambda}^{\omega}(r,0)\right](\beta, \mu) = - r^{2} +
\\
&& + \frac{1}{\beta V}\sum_{x\in\Lambda}\ln
{\mbox{Tr}}_{{\mathfrak{F}}_x} \exp \beta
\left[(\mu-\varepsilon_{x}^{\omega}-1)n_{x}- \lambda_{x}^{\omega}
n_{x}(n_{x}-1)+  r( \tilde{a}^{*}_{x} + \tilde{a}_{x})\right]
.\nonumber
\end{eqnarray}
Therefore, without source the \textit{maximizers} in
(\ref{max}) can be defined only up to a phase and their moduli
satisfy the equation:
\begin{equation}\label{max-1}
r = \frac{1}{2V}\sum_{x\in \Lambda}\left\langle  \tilde{a}_x +
\tilde{a}^*_x  \right\rangle_{H_{\Lambda}^{\omega}(r, 0)}=:
\xi_{\Lambda}^\omega(r),
\end{equation}
where
\begin{eqnarray}
&&\xi_{x}^\omega(r):= \left\langle  \tilde{a}_x + \tilde{a}^*_x
\right\rangle_{H_{\Lambda}^{\omega}(r, 0)} =\nonumber
\\&&=\frac{\mbox{Tr}_{{\mathfrak{F}}_x} \left\{(\tilde{a}_x +
\tilde{a}^*_x)\exp \beta
\left[(\mu-\varepsilon_{x}^{\omega}-1)n_{x}- \lambda_{x}^{\omega}
n_{x}(n_{x}-1)+ r (\tilde{a}^{*}_{x} +
\tilde{a}_{x})\right]\right\}}{\mbox{Tr}_{{\mathfrak{F}}_x} \exp
\beta \left[(\mu-\varepsilon_{x}^{\omega}-1)n_{x}-
\lambda_{x}^{\omega} n_{x}(n_{x}-1)+ r (\tilde{a}^{*}_{x} +
\tilde{a}_{x})\right]}.\label{rand-max}
\end{eqnarray}
When $r=0$, the approximating Hamiltonian (\ref{ham-appr}) is
invariant with respect to canonical gauge transformations
$\mathcal{U}_\varphi \tilde{a}_x \mathcal{U}_{\varphi}^* =
\tilde{a}_x e^{i\varphi}$ for any $\varphi$. This implies
$\xi_{x}^\omega(r = 0) = 0$. Hence, equation (\ref{max-1}) always
has a trivial solution $r = 0$ and , moreover,  by (\ref{bound-2})
any nontrivial solution $r_{\Lambda}^\omega \leq M$.

Finally, differentiating (\ref{rand-max}) with respect to $r$ we
obtain:
\begin{equation}\label{xi-prop}
0 \leq \partial_r \xi_{x}^\omega(r)\leq R,
\end{equation}
where, by the superstability (\ref{super-st}), the upper bound $R$
is finite uniformly in $\omega, r , x$. Hence, $-2M \leq
\partial_{r}\tilde{p}_{\Lambda,\,\rm{appr}}^{\omega}(\beta, \mu;
r)\leq 2RM$ for $r\in \left[0,M\right]$. By consequence the limit
(\ref{lim-appr-press}) implies the \textit{uniform} \textit{a.s.}
convergence of the sequence
$\left\{\tilde{p}_{\Lambda,\,\rm{appr}}^{\omega}(\beta, \mu; r)
\right\}_{\Lambda}$ for $r \in \left[0,M\right]$:
\begin{eqnarray}\label{lim-appr-press-1}
&& \tilde{p}_{\rm{appr}}(\beta, \mu; r) = a.s.-\lim_{\Lambda}
\tilde{p}_{\Lambda, \rm{appr}}^{\omega}(\beta, \mu; r)=
\\
&& = - r^2 + \beta^{-1}\mathbb{E}\left\{\ln
{{\mbox{Tr}}}_{{\mathfrak{F}}_x} \exp \beta
\left[(\mu-\varepsilon_{x}^{\omega}-1)n_{x}- \lambda_{x}^{\omega}
n_{x}(n_{x}-1)+ r (\tilde{a}^{*}_{x} + \tilde{a}_{x})\right]
\right\}, \nonumber
\end{eqnarray}
Therefore,
\begin{equation}\label{limsup=suplim}
a.s.-\lim_{\Lambda} \sup_{r\geq 0}\tilde{p}_{\Lambda,
\rm{appr}}^{\omega}(\beta, \mu; r)= \sup_{r\geq 0} \tilde{p}_{
\rm{appr}}(\beta, \mu; r).
\end{equation}
Together with (\ref{lim-1}) and (\ref{lim-appr-press}), the limit (\ref{limsup=suplim}) proves
the assertions (\ref{lim-press-1}) and (\ref{lim-press-2}) of the
theorem. \qquad \hfill $\square$
\begin{remark}\label{rem-1}
The function $\xi_{x}^\omega (r)$ is \textit{increasing} in $r$ by
virtue of (\ref{xi-prop}). Moreover, it has also been suggested
that for any $x \in {\mathbb{Z}}^{d}$ and $\omega \in \Omega$, the
function $ r \mapsto \xi_{x}^\omega (r)$ is concave, see
{\rm{\cite{BD}}} for discussion of this conjecture. This implies that the
\textit{nontrivial} solution of equation (\ref{max-1}) is
\textit{unique}. Notice that homogeneity and ergodicity of the
random field random field
$\left\{\varepsilon_{x}^{\omega}\right\}_{x\in{\mathbb{Z}}^{d}}$
implies the same for the random field
$\left\{\xi_{x}^\omega\right\}_{x\in{\mathbb{Z}}^{d}}$ defined by
(\ref{rand-max}). Therefore, equation (\ref{max-1}) in the
thermodynamic limit takes the form:
\begin{equation}\label{max-a.s-lim}
r = a.s.- \lim_{\Lambda} \xi_{\Lambda}^\omega(r) = \frac{1}{2}
\,\,\mathbb{E}\left(\xi_{x=0}^\omega(r)\right)=: f(r),
\end{equation}
expressing a \textit{self-averaging} property of the order
parameter $r$, see {\rm{\cite{PF}}}. Since the expectation in
(\ref{max-a.s-lim}) preserves convexity, solution of the limit
equation (\ref{max-a.s-lim}) should be also \textit{unique}.
Therefore, the sequence of maximizers
$\left\{r_{\Lambda}^{\omega}\right\}_\Lambda$ with $\mathbb{P}=1$
has a \textit{unique} accumulation point in the interval
$\left[0,M\right]$. Moreover, if $r_{\Lambda}^{\omega}$ is the
unique solution of equation (\ref{max-1}), then
\begin{equation}\label{solut-a.s.-lim}
a.s.- \lim_{\Lambda}r_{\Lambda}^{\omega}= r(\beta, \mu),
\end{equation}
where $r(\beta, \mu)$ denotes the unique solution of equation
(\ref{max-a.s-lim}).
\end{remark}
\textit{Proof}: Since $\lambda_{\rm min} >0$, by superstability
we get $r_{\Lambda}^{\omega}\leq M$, see (\ref{bound-2}), i.e.
\begin{equation}\label{limsup-liminf}
0\leq\lim_{\Lambda}\inf r_{\Lambda}^{\omega}\leq\lim_{\Lambda}\sup
r_{\Lambda}^{\omega}\leq M ,
\end{equation}
for any $\omega \in \Omega$. Now suppose that there exists
$\Omega_{>}$ with $\mathbb{P}(\Omega_{>})
>0$ and a subsequence $\left\{r_{\Lambda_n}^{\omega}
\right\}_{n\geq1}, \omega \in \Omega_{>}$ such that
\begin{equation}\label{*-sol}
\lim_{n\rightarrow\infty}r_{\Lambda_n}^{\omega}= r_{*}^{\omega} >
r(\beta, \mu),\,\,\,\omega \in \Omega_{>}.
\end{equation}
Then, by virtue of (\ref{max-1}), (\ref{xi-prop}),
(\ref{max-a.s-lim}) and (\ref{*-sol}) we get:
\begin{equation}\label{*-inequal}
\xi_{\Lambda_n}^\omega(r_{*}^{\omega}) - R
\left|r_{\Lambda_n}^{\omega}- r_{*}^{\omega}\right|\leq
r_{\Lambda_n}^{\omega}= \xi_{\Lambda_n}^\omega(r_{*}^{\omega} +
r_{\Lambda_n}^{\omega}- r_{*}^{\omega})\leq
\xi_{\Lambda_n}^\omega(r_{*}^{\omega}) + R
\left|r_{\Lambda_n}^{\omega}- r_{*}^{\omega}\right|.
\end{equation}
These estimates, together with the limit (\ref{*-sol}) and
a.s.-convergence of $\xi_{\Lambda_n}^\omega(r)$ to $f(r)$ for any
$r$ imply
\begin{equation}\label{*-inequal-lim}
r_{*}^{\omega}= f(r_{*}^{\omega})> r(\beta, \mu),
\end{equation}
for any $\omega \in \Omega_{>}$ with $\mathbb{P}(\Omega_{>}) >0$,
which is impossible by uniqueness of solution of
(\ref{max-a.s-lim}). Similarly one excludes the hypothesis
$r_{*}^{\omega} < r(\beta, \mu)$, which proves
(\ref{solut-a.s.-lim}). \qquad \hfill $\square$

\section{Limiting Hamiltonians}
\setcounter{equation}{0}
\renewcommand{\theequation}{\arabic{section}.\arabic{equation}}

\subsection{Limit of Hard-Core Bosons}

The \textit{hard-core} (h.c.) interaction in the Bose-Hubbard
model (\ref{Bose-Hubb}) corresponds to $\lambda = +\infty$, or
$\lambda_{\rm min}= +\infty$ for the IRH Bose-Hubbard model
(\ref{ham-1}). This  formally discards from the boson Fock space
$\mathfrak{F}_B(\Lambda)$ all vectors with \textit{more than one}
particle at one site.

Let $\Phi_0$ denote the \textit{vacuum vector} in
$\mathfrak{F}_B(\Lambda)$. Then the subspace
$\mathfrak{F}_{B}^{h.c.}(\Lambda)\subset \mathfrak{F}_B(\Lambda)$,
which corresponds to the hard-core restrictions, is spanned by the
orthonormal vectors
\begin{equation}\label{h.c.vectors}
\Phi_X = \prod_{x \in X} a_{x}^* \ \Phi_0 \ , \ \ \ \ X \subset \Lambda \ .
\end{equation}
Since the subspace $\mathfrak{F}_{B}^{h.c.}(\Lambda)$ is closed,
there is orthogonal projection $P_\Lambda$ onto
$\mathfrak{F}_{B}^{h.c.}(\Lambda)$ such that
\begin{equation}\label{projector}
\mathfrak{F}_{B}^{h.c.} (\Lambda) = P \ \mathfrak{F}_B (\Lambda) \
,
\end{equation}
and we get the representation
\begin{equation}\label{direct-sum}
\mathfrak{F}_B (\Lambda) = \mathfrak{F}_{B}^{h.c.} (\Lambda)
\oplus (\mathfrak{F}_{B}^{h.c.}(\Lambda))^{\bot} \ ,
\end{equation}
where the orthogonal compliment
$(\mathfrak{F}_{B}^{h.c.}(\Lambda))^{\bot}:= (I -
P)\mathfrak{F}_B(\Lambda)$.

Since our main Theorem \ref{main-1} is valid for any $\lambda_{\rm
min}> 0$ and the estimate (\ref{lim-1}) is uniform in
$\lambda_{x}^{\omega}$, we can extend this theorem to the
hard-core case by taking the limit $\lambda_{\rm min}\rightarrow +
\infty$.

For simplicity we consider the case of a sequence of non-random
identical and increasing positive $\{\lambda_{x}^\omega =
\lambda_s >0\}_{s=1}^{\infty}$ such that $\lambda_s \rightarrow
+\infty$.
\begin{lemma}\label{s-resolv-conv}
Let $\lambda_{s}\rightarrow + \infty$. Then for all $\zeta \in
\mathbb{C}: Im(\zeta) \neq 0$, and for any $\omega \in \Omega$ and
$\nu \in \mathbb{C}$ we have the strong resolvent convergence of
Hamiltonians (\ref{ham-nu}):
\begin{equation}\label{resolv-conv-1}
\lim_{\lambda_{s}\rightarrow + \infty}(H_{\Lambda}^{\omega}(s,\nu)
- \zeta I)^{-1} \Psi= P\left[T_{\Lambda} + \sum_{x
\in\Lambda}\varepsilon_{x}^{\omega}n_{x} - \sqrt{V}(
\overline{\nu} \hat{a}_0 + \nu \hat{a}_0 ^{*})- \zeta
I\right]^{-1}P \Psi \ , \ \Psi\in \mathfrak{F}_{B}(\Lambda) \ ,
\end{equation}
where
\begin{equation}\label{s-Ham}
H_{\Lambda}^{\omega}(s,\nu):= T_{\Lambda} + \lambda_{s} \sum_{x \in\Lambda}
n_{x}(n_{x}-1) + \sum_{x
\in\Lambda} \varepsilon_{x}^{\omega}n_{x} -
\sqrt{V}( \overline{\nu} \hat{a}_0 + \nu \hat{a}_0 ^{*}) \ .
\end{equation}
The same is true for approximating Hamiltonians
(\ref{ham-appr-nu}):
\begin{eqnarray}\label{resolv-conv-2}
&&\lim_{\lambda_{s}\rightarrow + \infty}(H_{\Lambda}^{\omega, appr}(s,z,\nu) - \zeta I)^{-1}\Psi = \\
\nonumber && P\left[ V\left|z\right|^2 - \sqrt{V}( \overline{z}
\hat{a}_0 + z \hat{a}_0 ^{*})+ \sum_{x \in
\Lambda}(\varepsilon_{x}^{\omega}+1) n_{x} - \sqrt{V}(
\overline{\nu} \hat{a}_0 + \nu \hat{a}_0 ^{*}) - \zeta I
\right]^{-1}P \Psi \,,
\end{eqnarray}
for any $z \in \mathbb{C}$ and $\Psi \in
\mathfrak{F}_{B}(\Lambda)$. Here
\begin{equation}\label{s-appr-Ham}
H_{\Lambda}^{\omega, appr}(s,z,\nu): = V\left|z\right|^2 - \sqrt{V}(\overline{z} \hat{a}_0 + z \hat{a}_0
^{*}) + N_{\Lambda} + \lambda_{s} \sum_{x \in\Lambda}
n_{x}(n_{x}-1) + \sum_{x
\in\Lambda} \varepsilon_{x}^{\omega}n_{x} -
\sqrt{V}( \overline{\nu} \hat{a}_0 + \nu \hat{a}_0 ^{*}) \ .
\end{equation}
\end{lemma}
\noindent \textit{Proof}: By estimate (\ref{super-st}) and
(\ref{s-Ham}) for $0<\lambda_{s}< \lambda_{s+1}$  we get:
\begin{equation}\label{sequence}
\frac{\lambda_{s}}{V}N_{\Lambda}^2  + (\varepsilon_{\rm min}
-\lambda_{s} - \left|\nu \right|^2) \ N_{\Lambda} - V \leq
H_{\Lambda}^{\omega}(s,\nu) \leq H_{\Lambda}^{\omega}(s+1,\nu) \ .
\end{equation}
So, for any $\omega \in \Omega$ and $\nu \in \mathbb{C}$
Hamiltonians (\ref{s-Ham}) form an increasing sequence of
self-adjoint operators, semi-bounded from below. Let
$\{h_{s}^{\omega}(\nu, \Lambda)[\Psi]:=
(\Psi,H_{\Lambda}^{\omega}(s,\nu)
\Psi)_{\mathfrak{F}_B(\Lambda)}\}_{s=1}^{\infty}$ be the
corresponding monotonic sequence of closed symmetric quadratic forms with domains
$ \ \mbox{dom}\ h_{s}^{\omega}(\nu, \Lambda)$. Put
\begin{equation}\label{Q-dom}
Q:= \bigcap_{s\geq1} \ \mbox{dom} \ h_{s}^{\omega}(\nu, \Lambda) \ ,
\end{equation}
and let $Q_0 = \overline{Q}$ be the closure of $Q$ in the Hilbert space $\mathfrak{F}_B(\Lambda)$.
Since for any $\omega \in \Omega$ and $\nu \in \mathbb{C}$
\begin{equation}\label{inf-lim}
\lim_{\lambda_{s}\rightarrow + \infty}(\Psi,H_{\Lambda}^{\omega}
(s,\nu)\Psi)_{\mathfrak{F}_B(\Lambda)} = + \infty \ , \ \ \Psi \in
(\mathfrak{F}_{B}^{h.c.}(\Lambda))^{\bot} \ \ ,
\end{equation}
one gets $Q_0 = \mathfrak{F}_{B}^{h.c.}(\Lambda)$ and
the strong resolvent convergence (\ref{resolv-conv-1}) of Hamiltonians, see e.g.
\cite{D}, Ch.4.4 or \cite{NZ}, Lemma 2.10. (Note that for hard cores the space
$\mathfrak{F}_{B}^{h.c.}(\Lambda)$ is finite-dimensional, which makes these arguments even simpler.)
The strong resolvent convergence (\ref{resolv-conv-1}) of Hamiltonians implies also
\begin{eqnarray}\label{s-res-conv-oper}
&&\lim_{\lambda_{s}\rightarrow + \infty}
(\Phi,H_{\Lambda}^{\omega}(s,\nu)\Phi)_{\mathfrak{F}_B (\Lambda)} = \\
\nonumber &&(\Phi,P [T_{\Lambda} + \sum_{x
\in\Lambda}\varepsilon_{x}^{\omega}n_{x} - \sqrt{V}(
\overline{\nu} \hat{a}_0 + \nu \hat{a}_0
^{*})]P\Phi)_{\mathfrak{F}_{B}^{h.c.}(\Lambda)} \ , \ \Phi \in  \mathfrak{F}_{B}^{h.c.}(\Lambda)\ .
\end{eqnarray}
The same line of reasoning leads to (\ref{resolv-conv-2}) for approximating
Hamiltonians. \qquad \hfill $\square$

By the Trotter approximating theorem \cite{RS I} the convergence
(\ref{resolv-conv-1}) and (\ref{resolv-conv-2}) yields the strong
convergence of the Gibbs semigroups:

\begin{corollary}\label{conv-gibbs-semig}
The following strong limits exist:
\begin{equation}\label{semig-1}
\rm{s}-\lim_{\lambda_{s}\rightarrow + \infty} e^{-\beta H_{\Lambda}^{\omega}(s,\nu)} =
e^{-\beta H_{h.c.,\Lambda}^{\omega}(\nu)} \ \ ,
\end{equation}
where
\begin{equation}\label{H-h.c.}
H_{h.c.,\Lambda}^{\omega}(\nu): = P [T_{\Lambda} + \sum_{x
\in\Lambda}\varepsilon_{x}^{\omega}n_{x} - \sqrt{V}(
\overline{\nu} \hat{a}_0 + \nu \hat{a}_0 ^{*})]P \ , \
\end{equation}
and similarly
\begin{equation}\label{semig-2}
\rm{s}-\lim_{\lambda_{s}\rightarrow + \infty} e^{-\beta
H_{\Lambda}^{\omega, appr}(s,z,\nu)} = e^{-\beta
H_{h.c.,\Lambda}^{\omega, appr}(z,\nu)} \ , \ \ \ {\rm{dom}}\
H_{h.c.,\Lambda}^{\omega}(\nu) = \mathfrak{F}_{B}^{h.c.}(\Lambda) \,  ,
\end{equation}
where
\begin{equation}\label{H-appr-h.c.}
H_{h.c.,\Lambda}^{\omega, appr}(z,\nu): =
P[ V\left|z\right|^2 - \sqrt{V}( \overline{z} \hat{a}_0 + z \hat{a}_0 ^{*})+
\sum_{x\in\Lambda}(\varepsilon_{x}^{\omega}+1) n_{x} -
\sqrt{V}( \overline{\nu} \hat{a}_0 + \nu \hat{a}_0 ^{*})]P  \,,
\end{equation}
with $\ {\rm{dom}}\ H_{h.c.,\Lambda}^{\omega, appr}(z,\nu) =
\mathfrak{F}_{B}^{h.c.}(\Lambda)$.
\end{corollary}
Since $\left\{e^{-\beta (H_{\Lambda}^{\omega}(s,\nu) - \mu N_\Lambda)}\right\}_{s\geq 1}$ is a sequence of
trace-class operators from
$\mathcal{C}_{1}(\mathfrak{F}_{B}(\Lambda))$ monotonously
decreasing to the trace-class operator
\begin{equation*}
e^{-\beta (H_{h.c.,\Lambda}^{\omega}(\nu) - \mu N_\Lambda)} \in
\mathcal{C}_{1}(\mathfrak{F}_{B}^{h.c.}(\Lambda)) \ ,
\end{equation*}
the convergence (\ref{semig-1}) can be lifted to the trace-norm
topology, see \cite{Z}. The same is true for (\ref{semig-2}). It
then follows that the pressures also converge:
\begin{lemma} \label{pressure-lim}
\begin{equation}\label{hard-core-press-1}
\lim_{\lambda_{s}\rightarrow + \infty} p[H_{\Lambda}^{\omega}(s,\nu)]=
p[H_{h.c.,\Lambda}^{\omega}(\nu)] \ ,
\end{equation}
\begin{equation}\label{hard-core-press-2}
\lim_{\lambda_{s}\rightarrow + \infty} p[H_{\Lambda}^{\omega}(s,z,\nu)]=
p[H_{h.c.,\Lambda}^{\omega, appr}(z,\nu)] \ .
\end{equation}
\end{lemma}
Since the estimate (\ref{lim-1}) is uniform in $\lambda \geq
\lambda_{\rm min}>0$, we can take the limit
$\lambda_{s}\rightarrow + \infty$  to obtain
\begin{equation}\label{h.c.-Bog}
0 \leq p\left[H_{h.c.,\Lambda}^{\omega}(\nu)\right] -
p\left[H_{h.c.,\Lambda}^{\omega, appr}(z_{\Lambda,\omega}(\nu), \nu)
\right]\leq O(1/V) \ ,
\end{equation}
for all $\omega\in\Omega$, any $\beta>0$, $\mu\in \mathbb{R}^1$
and $\left|\nu\right|< C_0$. Then, by the same line of reasoning
as after (\ref{lim-1}) in Theorem \ref{main-1}, we obtain the
thermodynamic limit of the pressure for the hard-core bosons:
\begin{corollary}\label{h.c.-var}
The pressure of the Infinite-Range-Hopping hard-core Bose-Hubbard
model with randomness is given by
\begin{eqnarray}\label{press-h.c.}
&& p_{h.c.}(\beta,\mu)= \\
\nonumber &&\sup_{r\geq0}\left\{-r^{2}+
\beta^{-1}\mathbb{E}\{\ln \mbox{{\rm{Tr}}}_{(\mathfrak{F}_{B}^{h.c.})_x}
\exp(\beta P\left[(\mu-\varepsilon_{x}^{\omega}-1)n_{x}+ r (a_{x}^* +
a_{x})\right]P)\}\right\} \ ,
\end{eqnarray}
cf. expression (\ref{lim-press-2}) for finite $\lambda$.
\end{corollary}
\begin{remark}\label{h.-c.-oper} To calculate the {\rm{Tr}} over
$\mathfrak{F}_{B}^{h.c.}$ note that
the boson creation and annihilation operators are quite different
from operators {\rm{:}} $c_{x}^*:=Pa_{x}^*P$, $c_{x}:=Pa_{x}P$ restricted
to ${\rm{dom}} \ c_{x}^* = {\rm{dom}} \ c_{x} =
\mathfrak{F}_{B}^{h.c.}$, which occur in (\ref{press-h.c.}). The
major difference consists in their commutation relations:
\begin{equation}\label{CAR}
[c_{x}, c_{y}^*]=0 \,, \ (x\neq y) \ , \ \ \ (c_{x})^2= (c_{x}^*)^2 =0 \ , \ \
\ c_{x} c_{x}^* + c_{x}^* c_{x} = I \ .
\end{equation}
Taking the $XY$ representation of relations (\ref{CAR}) {\rm{:}}
\begin{equation*}
c_x  = \left(
               \begin{array}{cc}
                 0 & 0 \\
                 1 & 0 \\
               \end{array}
             \right) \ , \ \ \ \ c_{x}^* = \left(
               \begin{array}{cc}
                 0 & 1 \\
                 0 & 0 \\
               \end{array}
             \right) \ ,
\end{equation*}
(\ref{press-h.c.}) gives explicitly
\begin{eqnarray}\label{press-h.c.-1}
&& p_{h.c.}(\beta,\mu)= \\
\nonumber && \sup_{r\geq0}\left\{-r^{2}+
\mathbb{E}\left\{\frac{1}{2}(\mu-\varepsilon_{x}^{\omega}-1)+
\beta^{-1}\ln \left[2 \cosh \left(\frac{1}{2}\beta
\sqrt{(\mu-\varepsilon_{x}^{\omega}-1)^2 + 4r^2}\right)\right]\right\}\right\} \ ,
\end{eqnarray}
the grand-canonical pressure for the random IRH hard-core
Bose-Hubbard model.
\end{remark}

\subsection{Limit of Perfect Bosons}

The limit $\lambda \to 0$ is more delicate. For simplicity, below we assume that $\varepsilon_{\rm min} =
0$. Then Hamiltonian (\ref{ham-1}) for perfect bosons  $\lambda_{x}^\omega = 0$ is non-negative, i.e.
the corresponding pressure exists in a finite volume only for \textit{negative} chemical potentials.
There is an analogue of Lemma~\ref{s-resolv-conv}, if we subtract from this Hamiltonian a term $\mu
N_\Lambda$ with $\mu < 0$ and assume $\nu$ small enough:

\begin{lemma}\label{s-resolv-conv-2}
Assume that $\varepsilon_{\rm min} =
0$ and let $\lambda_{s}\searrow 0$. Then for $\mu < 0$, for all $\zeta \in \mathbb{C}:
Im(\zeta) \neq 0$, and for any $\omega \in \Omega$, we have the strong resolvent convergence
of Hamiltonians (\ref{ham-nu}):
\begin{eqnarray}\label{resolv-conv-3}
&&\lim_{\lambda_{s}\searrow 0}(H_{\Lambda}^{\omega}(s,\nu) - \mu
N_\Lambda - \zeta I)^{-1} \Psi =  \\
&&\{T_{\Lambda} + \sum_{x
\in\Lambda} (\varepsilon_{x}^{\omega}-\mu) n_{x} - \sqrt{V}(
\overline{\nu} \hat{a}_0 + \nu \hat{a}_0 ^{*})- \zeta
I \}^{-1} \Psi \ , \ \Psi\in \mathfrak{F}_{B}(\Lambda) \ , \nonumber
\end{eqnarray}
for $\nu \in \mathbb{C}$, if $|\nu|^2 < |\mu|$.
The same is true for approximating Hamiltonians
(\ref{ham-appr-nu}):
\begin{eqnarray}\label{resolv-conv-4}
&&\lim_{\lambda_{s}\searrow 0} (H_{\Lambda}^{\omega,
appr}(s,z,\nu) - \mu N_\Lambda - \zeta I)^{-1} \Psi = \\ \nonumber
&& \{V\left|z\right|^2 - \sqrt{V}( \overline{z} \hat{a}_0 + z
\hat{a}_0 ^{*})+ \sum_{x \in\Lambda}
(\varepsilon_{x}^{\omega}+1-\mu) n_{x} - \sqrt{V}( \overline{\nu}
\hat{a}_0 + \nu \hat{a}_0 ^{*}) - \zeta I\}^{-1} \Psi \,,
\end{eqnarray}
for any $z \in \mathbb{C}$, $\zeta \in \mathbb{C}: Im(\zeta) \neq 0$ and $\Psi \in \mathfrak{F}_{B}(\Lambda)$.
\end{lemma}

\noindent \textit{Proof}: The bound (\ref{super-st}) now yields:
\begin{equation}\label{lower-bnd}
H_\Lambda^\omega (s,\nu,\mu):= H_\Lambda^\omega (s,\nu) - \mu N_\Lambda \geq (-\mu - |\nu|^2) N_\Lambda -
V \, ,
\end{equation}
so that for $|\nu|^2 + \mu < 0$, the operators
$\left\{H_\Lambda^\omega (s,\nu,\mu)\right\}_{s\geq1}$ are positive. As in
Lemma~\ref{s-resolv-conv}, for these operators we define the corresponding closed symmetric quadratic
forms by
$\{h_{s}^{\omega}(\nu, \mu, \Lambda)[\Psi]:=
(\Psi,H_{\Lambda}^{\omega}(s,\nu, \mu)
\Psi)_{\mathfrak{F}_B(\Lambda)}\}_{s=1}^{\infty}$.
Note that they are monotonously decreasing and bounded from below,
which implies that for any  $\omega \in \Omega$, $\nu \in \mathbb{C}$ and $\Lambda$
the operators $\left\{H_\Lambda^\omega (s,\nu,\mu)\right\}_{s\geq1}$ converge in
the strong resolvent sense, see e.g. \cite{K}, Ch.VIII, to a positive self-adjoint
operator $H_{\Lambda, 0}^\omega (\nu,\mu)$.
Let us define the symmetric form
\begin{equation}\label{lim-form}
h_{\infty}^{\omega}[\Phi] = \lim_{s\rightarrow\infty} h_{s}^{\omega}[\Phi] \, ,
\end{equation}
with domain
\begin{equation*}
\mbox{dom}\,(h_\infty ^{\omega}) = \bigcup_{s \geq 1} \ \mbox{dom}\,(h_s ^{\omega}) \, .
\end{equation*}
It is known, \cite{K}  Ch.VIII, that if the form (\ref{lim-form}) is closable, then operator
$H_{\Lambda, 0}^\omega (\nu,\mu)$ is associated with the closure $\tilde{h}_{\infty}^{\omega}$.
By explicit expression of $h_{s}^{\omega}(\nu, \mu, \Lambda)$
one gets that the limit form (\ref{lim-form}) is closable (and even closed), since it is
associated with the self-adjoint
operator $H_\Lambda^\omega (s=\infty,\nu,\mu)$. Then the operator $H_{\Lambda, 0}^\omega (\nu,\mu)$
associated with the
closure $\tilde{h}_{\infty}^{\omega}$ of (\ref{lim-form}) simply coincides with
$H_\Lambda^\omega (s=\infty,\nu,\mu)$:
\begin{equation*}
\tilde{h}_\infty ^{\omega}[\Phi] = ( \Phi\,,\, [
T_\Lambda + \sum_{x \in \Lambda} (\varepsilon_x^\omega - \mu) n_x -
\sqrt{V} (\overline{\nu} \hat{a}_0 + \nu \hat{a}_0 ^{*})
] \ \Phi ) \ ,
\end{equation*}
that proves (\ref{resolv-conv-3}).

A similar argument applies for the approximating Hamiltonians (\ref{ham-appr-nu}).
But, in contrast to the case of
sources $|\nu|^2 < |\mu|$, that we can choose as small as we want to apply  the main
Theorem \ref{main-1},
the value of $z$ will be defined by variational principle (\ref{lim-press-2})
with $\lambda_{x}^\omega \geq 0$. Now
the semi-boundedness of $\left\{H_{\Lambda}^{\omega, appr}(s,z,\nu)\right\}_{s\geq1} \ $
from below follows from the
estimate
\begin{equation}\label{est-appr-below}
\sum_{x \in\Lambda}
(\varepsilon_{x}^{\omega}+1-\mu) n_{x} - \sqrt{V}( (\overline{\nu} + \overline{z})
\hat{a}_0 + (\nu + z) \hat{a}_0 ^{*}) \geq - \ V \ \frac{|\nu + z|^2}{1 - \mu} \ .
\end{equation}
The rest of the arguments is identical to those for the operators  (\ref{lower-bnd}),
or equivalently for the  sequence
$\left\{H_{\Lambda}^{\omega}(s,\nu)\right\}_{s\geq1} \ $, and goes through verbatim to
give the proof of the limit
(\ref{resolv-conv-4}) with $H_{\Lambda}^{\omega, appr}(s=\infty,z,\nu) :=
H_{\Lambda, 0}^{\omega, appr}(z,\nu) $.
\qquad \hfill $\square$

\begin{corollary}\label{pressure-lambda-zero}
In a full analogy with Corollary \ref{conv-gibbs-semig}  and Lemma \ref{pressure-lim},  the
Trotter approximation
theorem and the monotonicity of the operator families
$\left\{H_{\Lambda}^{\omega}(s,\nu)\right\}_{s\geq1} \ $,
$\left\{H_{\Lambda}^{\omega, appr}(s,z,\nu)\right\}_{s\geq1} \ $ yield
\begin{equation}\label{zero-press-1}
\lim_{\lambda_{s}\rightarrow 0} p[H_{\Lambda}^{\omega}(s,\nu)]=
p[H_{\Lambda, 0}^{\omega}(\nu)] \ ,
\end{equation}
\begin{equation}\label{zero-press-2}
\lim_{\lambda_{s}\rightarrow 0} p[H_{\Lambda}^{\omega, appr}(s,z,\nu)]=
p[H_{\Lambda, 0}^{\omega, appr}(z,\nu)] \ .
\end{equation}
Notice that, similarly to the Weakly Imperfect Bose-Gas {\rm{\cite{ZB}}}, the
estimate (\ref{lim-1}) for $\mu < 0 $ is still uniform in $\lambda \geq 0$.
Therefore, we can take there the limit $\lambda_{s}\rightarrow 0$ to obtain
\begin{equation}\label{0-Bog}
0 \leq p\left[H_{\Lambda, 0}^{\omega}(\nu)\right] -
p\left[H_{\Lambda, 0}^{\omega, appr}(z_{\Lambda,\omega}(\nu), \nu)
\right]\leq O(1/V) \ ,
\end{equation}
for all $\omega\in\Omega$, any $\beta>0$ and $\left|\nu\right|^2< -
\mu$. Then, following the same line of reasoning as after (\ref{lim-1}) in
Theorem \ref{main-1}, we obtain the thermodynamic limit of the
pressure for the perfect bosons:
\begin{eqnarray}\label{press-0}
&& p_{0}(\beta,\mu < 0)= \\
\nonumber &&\sup_{r\geq0}\left\{-r^{2}+ \beta^{-1}\mathbb{E}\{\ln
\mbox{{\rm{Tr}}}_{(\mathfrak{F}_{B})_x} \exp(\beta
\left[(\mu-\varepsilon_{x}^{\omega}-1)n_{x}+ r (a_{x}^* +
a_{x})\right])\}\right\} \ ,
\end{eqnarray}
cf. expression (\ref{lim-press-2}) for finite $\lambda$, where all values of $\mu$ are allowed.
Since we put $\varepsilon_{\rm min} = 0$, the variational principle in (\ref{press-0}) implies:
\begin{eqnarray}\label{press-0-negat-mu}
&&p_{0}(\beta,\mu < 0)= \beta^{-1}\mathbb{E}\{\ln
\mbox{{\rm{Tr}}}_{(\mathfrak{F}_{B})_x} \exp(\beta
\left[(\mu-\varepsilon_{x}^{\omega}-1)n_{x}\right])\} = \\
&&\beta^{-1}\mathbb{E}\left\{\ln \left[1 - \exp \{\beta
(\mu-\varepsilon_{x}^{\omega}-1)\}\right]^{-1}\right\} \ . \nonumber
\end{eqnarray}
The convexity of  $\left\{p\left[H_{\Lambda, 0}^{\omega}(\nu=0)\right]\right\}_{\Lambda}$ and
the thermodynamic limit $p_{0}(\beta,\mu)$  as the functions of $\mu < 0$, together with the Griffith lemma,
see e.g. {\rm{\cite{ZB}}}, yield the convergence of derivative with respect of $\mu$, i.e. the formula for
the \textit{total} particle density:
\begin{equation}\label{dens-0}
\rho(\beta, \mu <0) = \mathbb{E} \left[ \frac{1}{e^{\beta (1+\varepsilon^\omega -\mu)}-1}
\right].
\end{equation}
\end{corollary}
\begin{remark}\label{perfect-bg-mu-0}
As usually in the case of the perfect boson gas one recovers the value of thermodynamic parameters
at extreme point  $\mu=0$ by continuation: $\mu \rightarrow -0$:
\begin{eqnarray}
&& p_{0}(\beta,\mu= 0): =  \beta^{-1}\mathbb{E}\left\{\ln \left[1 - \exp \{\beta
(-\varepsilon_{x}^{\omega}-1)\}\right]^{-1}\right\} \ , \label{press-perf-0} \\
&&\rho(\beta, \mu= 0): = \mathbb{E} \left[ \frac{1}{e^{\beta (1+\varepsilon^\omega)}-1}\right]
\ . \label{dens-perf-0}
\end{eqnarray}
In particular by (\ref{dens-perf-0}) it gets clear that the gap {\rm{($= 1$)}} in the one-particle
spectrum of
the perfect boson gas $T_\Lambda$ and $\varepsilon_{\rm min} = 0$ imply that the critical density
\begin{equation}\label{cr-dens-perf}
\rho_c(\beta):= \sup_{\mu < 0} \rho(\beta, \mu) = \rho(\beta, \mu= 0)
\end{equation}
is finite, cf. (\ref{gap}) and (\ref{cr-dens-free-ir}). This opens a room for the zero-mode Bose
condensation
in the case of the random potential $\left\{\varepsilon_{x}^\omega\right\}_x$.
\end{remark}


\section{Phase Diagram}

\setcounter{equation}{0} \renewcommand{\theequation}{\arabic{section}.%
\arabic{equation}} 

Here we analyse only  the case, when $\varepsilon_{x}^\omega$ is
random, but the interaction couplings $\lambda_{x}^\omega = \lambda
\geq 0$ are fixed.

To proceed we recall first the formulae determining the critical
temperature $\beta_c(\rho,\lambda)^{-1}$ for the \textit{nonrandom}
case $\varepsilon_{x}^\omega =0 $. To this end we define, cf (\ref{appr-press-1}),
\begin{equation}\label{press-tilda}
\tilde p(\beta, \mu, \lambda; r):= \frac{1}{\beta} \ln
{\mbox{Tr}}_{{\mathcal{H}}} \exp (-\beta \left[h_{n} (\mu, \lambda)
- r({a}^{*} + {a})\right]) \ ,
\end{equation}
where
\begin{equation}\label{h}
h_{n} (\mu, \lambda):= (1-\mu)n + \lambda n(n-1) \ .
\end{equation}
Due to \cite{BD} it is known that the critical temperature (and the
critical chemical potential $\mu_c (\rho, \lambda)$) are defined, as
functions of the total particle density $\rho$, by two equations:
\begin{equation}\label{crit-temp}
\tilde p''(\beta, \mu, \lambda; 0) = 2 \ , \ \ \rho = \frac{1}
{Z_0(\beta,\mu,\lambda)} \sum_{n=1}^\infty n\,e^{-\beta
h_n(\mu,\lambda)} \ .
\end{equation}
Here
\begin{equation}\label{ptilde}
{\tilde p}''(\beta,\mu,\lambda;0) = \frac{2}{Z_0(\beta,\mu,\lambda)}
\sum_{n=1}^\infty n \frac{e^{-\beta h_n(\mu,\lambda)} - e^{-\beta
h_{n-1}(\mu,\lambda)}}{h_{n-1}(\mu,\lambda) - h_n(\mu,\lambda)} \ .
\end{equation}
and
\begin{equation*}
Z_0(\beta,\mu,\lambda) = \sum_{n=0}^\infty e^{-\beta
h_n(\mu,\lambda)} \ .
\end{equation*}

If $\varepsilon_{x}^\omega \neq 0 $ and $\lambda > 0$, then by the
main Theorem \ref{main-1} (see (\ref{lim-press-2}),
(\ref{lim-appr-press-1}) and  (\ref{h})) to obtain the equations for
the critical temperature and the critical chemical potential we have
to replace $\mu$ in (\ref{crit-temp}) by $\mu -
\varepsilon_x^\omega$ and to average over $\varepsilon_x^\omega$.
This gives, instead of (\ref{crit-temp}), the (\textit{gap}) equation:
\begin{equation}\label{betacrit}
\mathbb{E} \left[ {\tilde
p}''(\beta,\mu-\varepsilon^\omega,\lambda;0) \right] = 2 \ ,
\end{equation}
and equation for density:
\begin{equation}\label{rhocrit}
\rho = \mathbb{E} \left[\frac{1}
{Z_0(\beta,\mu-\varepsilon^\omega,\lambda)} \sum_{n=1}^\infty
n\,e^{-\beta h_n(\mu-\varepsilon^\omega,\, \lambda)} \right].
\end{equation}

The case of $\lambda=0$ is more subtle, and we begin with it the next subsection.

\subsection{Perfect bosons: $\lambda = 0$}

Without loss of generality, we can assume that the random $\varepsilon^\omega$ takes
values in the interval $[0,\varepsilon]$. In that case the maximal allowed
value for $\mu$ (i.e. the \textit{critical value}) is still $\mu_c =0$, and the critical inverse temperature
$\beta_c : = \beta_c(\rho,\lambda =0)$ is given (see (\ref{dens-perf-0}), (\ref{cr-dens-perf}))  by:
\begin{equation}\label{crit-free}
\rho = \mathbb{E} \left[ \frac{1}{e^{\beta_c (1+\varepsilon^\omega)}-1}
\right].
\end{equation}

Remark that, \textit{irrespective} of the distribution of
$\varepsilon^\omega$, the equation (\ref{crit-free}) implies that
the resulting $\beta_c$ is \textit{lower} than $\ln \left(1 +
\frac{1}{\rho} \right)$, which corresponds to the nonrandom case
$\varepsilon_{x}^\omega = 0 $, i.e. \textit{disorder enhances}
Bose-Einstein condensation. We shall see (Sect.4.3.3) that this is \textit{no longer
true} when $\lambda > 0$, and even that the \textit{opposite} holds, if
$\lambda$ is small enough!

Notice that formula (\ref{crit-free}) is in agreement with the
general expression found in \cite{LPZ}:
\begin{equation}\label{beta-crit-free}
\rho = \int \frac{d {\bar {\cal
N}}(E)}{e^{\beta_c E}-1},
\end{equation} where ${\bar {\cal N}}(E)$ is the \textit{integrated} density
of states given by
\begin{equation}\label{int-dens-lim}
 {\bar {\cal N}}(E) = {\rm{a.s.-}}\lim_{V \to\infty}
\frac{1}{V} \#\{i:\,E_{i}^\omega \leq E\}.
\end{equation}
Here $\left\{E_{i}^\omega\right\}_{i\geq1}$ are the eigenvalues of the
one-particle Hamiltonian with a random potential $\left\{\varepsilon_{x}^\omega\right\}_{x\in\Lambda}$:
\begin{equation}\label{ham-free-rand}
(h_{\Lambda}^\omega u)(x):= (t_{\Lambda}u)(x) + \sum_{x \in \Lambda} \varepsilon_{x}^\omega u(x) \ , \
x\in\Lambda ,\ \ u\in\mathfrak{h}(\Lambda) ,
\end{equation}
for \textit{i.r.} kinetic-energy hopping, see
(\ref{one-kinetic-energy}), (\ref{hopp-inf-rang}), and $\# \{i:\,E_{i}^\omega \leq E \}$ counting
the number of the corresponding eigenfunctions (including the \textit{multiplicity}). It is known
that for any \textit{ergodic} random potential $\left\{\varepsilon_{x}^\omega\right\}_{x\in\Lambda}$,
the limit (\ref{int-dens-lim}) exists \textit{almost surely} (a.s.) and that it is \textit{non-random},
see e.g.\cite{PF}.
A contact between formulae (\ref{crit-free}) and (\ref{beta-crit-free}) gives the following
\begin{lemma} The integrated density
of states is equal to
\begin{equation}\label{DOS}
{\bar {\cal N}}(E) = \mathbb{P} \left[ \varepsilon^\omega \leq E-1
\right] = \mathbb{E} \ [\theta (E - (1 + \varepsilon^\omega))] \ .
\end{equation}
\end{lemma}
\textit{Proof}: For simplicity we consider the case of a \textit{Bernoulli} random
potential $\left\{\varepsilon_{x}^\omega\right\}_{x\in \Lambda}$
such that $\varepsilon_{x}^\omega = \varepsilon$ with probability
$p$ and $\varepsilon_{x}^\omega = 0$ with probability $1-p$. (The
proof of the general case is similar, but slightly more
complicated.) In this special case, the right-hand side of
(\ref{DOS}) equals
\begin{equation}\label{Prob}
\mathbb{P} \left[ \varepsilon^\omega \leq E-1 \right] = \left\{
\begin{array}{ll} 1 & \mbox{if } E \geq 1+\varepsilon, \\ 1-p &\mbox{if } 1
\leq E < 1+\varepsilon, \\ 0 & \mbox{if } E < 1.
\end{array} \right.
\end{equation}

Clearly, all eigenvalues $\{E_{i}^\omega\}_{i \geq 1}$  of the
Hamiltonian (\ref{ham-free-rand}) belong to the interval $[1,
1+\varepsilon]$. Since $\dim(\mathfrak{h}(\Lambda))= V$, one gets
${\bar {\cal N}}(E) = 1$, if $E \geq 1+\varepsilon$. Similarly,
${\bar {\cal N}}(E) = 0$, if $E < 1$.

Now suppose that $E \in [1,1+\varepsilon)$. Since
$\left\{\varepsilon_{x}^\omega\right\}_{x\in \Lambda}$ is the
Bernoulli random field, for given $\delta > 0$, there
exists $c > 0$ such that with probability $Pr > 1-\delta$ the number
of sites $x\in \Lambda$ with $\varepsilon_{x}^{\omega} =
\varepsilon$ is in the interval $(p V - c \sqrt{V}, pV + c
\sqrt{V})$. Given a configuration for which this is the case, let
$\Lambda_{\varepsilon} \subset \Lambda$ be the set where
$\varepsilon_{x}^{\omega} = \varepsilon$. Consider the states $\phi \in
\mathfrak{h}(\Lambda)$ such that $\phi(x) = 0$, if $x \notin
\Lambda_{\varepsilon}$ and $\sum_{x \in \Lambda} \phi(x) = 0$. Then
\begin{equation*}
(h_{\Lambda}^\omega \phi)(x) = \frac{1}{V} \sum_{y=1}^V (\phi(x) - \phi(y)) +
\varepsilon^\omega_x \phi(x) =  (\varepsilon + 1) \phi(x) \ , \ \ x\in \Lambda_{\varepsilon} .
\end{equation*}
The space of such eigenfunctions $\phi$ has dimension $|\Lambda_{\varepsilon}|-1$, so that
\begin{equation*}
\#\{E_{i}^\omega >
E\} \geq (|\Lambda_{\varepsilon}|-1).
\end{equation*}
Since $(\#\{E_{i}^\omega \leq E\}) + (\#\{E_{i}^\omega >
E\}) = V$, for $V \to \infty$ we get
\begin{equation*}
{\bar {\cal
N}}(E) \leq 1-p  .
\end{equation*}
Similarly, considering the eigenfunctions with supports concentrated
on $\Lambda_{\varepsilon}^c = \Lambda \setminus \Lambda_{\varepsilon} $ we obtain
\begin{equation*}
{\bar {\cal N}}(E) \geq 1-p .
\end{equation*}
Together with (\ref{Prob}) these estimates give the proof of (\ref{DOS}).
\qquad \hfill $\square$

The relations (\ref{DOS}) show that the formulae (\ref{crit-free}) and (\ref{beta-crit-free})
are equivalent.
For details of a general statement see e.g. \cite{PF} Ch.II.5 .

\subsection{Discrete random potential and $\lambda > 0$}

We now consider the case with interaction $\lambda > 0$, and first assume that the
probability distribution of $\varepsilon^\omega_x$ is \textit{discrete}.

A particularly simple case corresponds to the \textit{hard-core}
boson limit $\lambda = +\infty$, see Section 3. Then by
(\ref{press-h.c.-1}) the equations for the critical value of the
inverse temperature  $\beta_c := \beta_c(\rho)= \beta_c(\rho,\lambda
= +\infty)$ for a given density $\rho$, reduce to the system:
\begin{equation}\label{betacrit1}
\mathbb{E} \left[ \frac{\tanh \beta (\mu - \varepsilon^\omega
-1)/2}{\mu - \varepsilon^\omega-1} \right] = 1
\end{equation}
and
\begin{equation}\label{rhocrit1}
\rho = \frac{1}{2} + \frac{1}{2} \mathbb{E} \left[
\tanh \frac{1}{2} \beta (\mu - \varepsilon^\omega - 1) \right] \ .
\end{equation}
The last equation (\ref{rhocrit1}) implies that for the hard-core
interaction the total particle density has the estimate:  $\rho \leq
1$.

\subsubsection{Bernoulli random potential in the hard-core limit $\lambda = + \infty$.}

A special case of a discrete distribution is the Bernoulli
distribution, where $\varepsilon^\omega_x = \varepsilon$ with
probability $p$ and $\varepsilon^\omega_x=0$ with probability
$1-p$. We first consider the case $\lambda = +\infty$. The
equations (\ref{betacrit1}) and (\ref{rhocrit1}) then read,
\begin{equation} F_{p,\varepsilon}(\beta=\beta_c , \mu):=
p \frac{\tanh \frac{1}{2} \beta_c (\mu -
\varepsilon-1)}{\mu - \varepsilon-1} + (1-p) \frac{\tanh \frac{1}{2}
\beta_c (\mu -1)} {\mu-1} = 1 \label{bc_bin}
\end{equation}
and
\begin{equation} G_{p,\varepsilon}(\beta=\beta_c , \mu):=
\frac{1}{2} + \frac{1}{2} \left[ p \tanh \frac{1}{2}
\beta_c (\mu - \varepsilon - 1) + (1-p) \tanh \frac{1}{2} \beta_c
(\mu-1) \right] = \rho. \label{rho_bin}
\end{equation}

Here a \textit{new phenomenon} occurs for density $\rho = 1-p$. To
see this, we consider first a particular case of $p = {1}/{2}$. Then
$\rho={1}/{2}$, and by (\ref{rho_bin}) we obtain, that the only
possible solution for the corresponding chemical potential is
$\mu(\rho= 1/2):= \mu(\rho={1}/{2}, \lambda = + \infty) = 1 +
\varepsilon /2$. Inserting this value of $\mu$ into (\ref{bc_bin})
we get for the critical temperature:
\begin{equation*}
\tanh \frac{\beta_c \varepsilon}{4} = \frac{1}{2} \varepsilon \ .
\end{equation*}
This equation obviously has \textit{no solution} for $\varepsilon
\geq 2$. Therefore, there is  \textit{no} Bose-Einstein condensation
for Bernoulli random potential, if  $ p= \rho = {1}/{2}$, and
$\varepsilon$ is greater than some  \textit{critical} value:
$\varepsilon_{cr} = 2$.

One can check that the same phenomenon occurs for $p \neq {1}/{2}$  and for densities  $\rho = 1-p$,
if $\varepsilon$ is \textit{large} enough, but now the reasoning is more
delicate. First of all, by (\ref{bc_bin}) and $\tanh u \leq u$ we see that in any case there is a
\textit{lower bound} on the inverse critical temperature:
\begin{equation}
\beta_c \geq 2.
\label{betag2}
\end{equation}
Now assume that $p < 1/2$, i.e. $\rho > 1/2$. From (\ref{rho_bin}) it then follows
that for any $\varepsilon$ one has
\begin{equation}\label{estim-mu-1}
0 < \mu -1 - \frac{1}{2}\varepsilon  \ .
\end{equation}
Indeed, if we suppose that $0 \leq \mu - 1 \leq \varepsilon/2 $, then $\tanh \frac{1}{2}
\beta_c (\mu-1) \leq \tanh \frac{1}{2} \beta_c
(1+\varepsilon-\mu)$ and hence, by (\ref{rho_bin}), we get
\begin{eqnarray*}
\nonumber 2\rho -1 &=& \lefteqn{p \tanh \frac{1}{2}
\beta_c (\mu-\varepsilon-1) + (1-p)
\tanh \frac{1}{2} \beta_c (\mu-1) } \\
&\leq& (1-2p) \tanh \frac{1}{2} \beta_c (\varepsilon+1-\mu) < 1-2p
\ ,
\end{eqnarray*}
contradicting our assumption $\rho = 1-p$, if $\beta_c$ exists and is finite.

Now notice that (\ref{rho_bin}) with $\rho = 1-p$ is equivalent to
\begin{equation}
\frac{1-\tanh \frac{1}{2} \beta_c(\varepsilon+1-\mu)}{1 -\tanh
\frac{1}{2} \beta_c (\mu-1)} = \frac{1-p}{p} \ . \label{crden}
\end{equation}
The left-hand side of (\ref{crden}) can be estimated from below as
\begin{equation*}
\frac{1-\tanh \frac{1}{2}
\beta_c(\varepsilon+1-\mu)}{1 -\tanh \frac{1}{2} \beta_c (\mu-1)} =
\frac{e^{\beta_c (\mu - 1 - \varepsilon/2)} +
e^{- \beta_c \varepsilon/2}}{e^{-\beta_c (\mu - 1 -
\varepsilon/2)} + e^{- \beta_c \varepsilon/2}} >
e^{\beta_c (\mu-1-\varepsilon/2)} \ .
\end{equation*}
Together with (\ref{betag2}) this yield an upper bound for (\ref{estim-mu-1}):
\begin{equation}\label{upperbound}
0 < \mu-1-\frac{1}{2} \varepsilon < \frac{1}{\beta_c} \ln
\frac{1-p}{p} \leq \frac{1}{2} \ln \frac{1-p}{p} < \frac{1-2p}{2p} \ .
\end{equation}
But (\ref{upperbound}) implies that (\ref{bc_bin}) has \textit{no} solution $\beta_c$, since for
\textit{large} $\varepsilon$ we obtain
\begin{eqnarray}\label{estim-large-eps}
&& p \frac{\tanh \frac{1}{2} \beta_c (\mu - \varepsilon-1)}{\mu -
\varepsilon-1} + (1-p) \frac{\tanh \frac{1}{2} \beta_c (\mu -1)}
{\mu-1} <  \\
&& \nonumber \frac{p}{\varepsilon + 1 - \mu} + \frac{1-p}{\mu -1} \ < \
\frac{p}{\varepsilon/2 - (1-2p)/2p} + \frac{1-p}{\varepsilon/2} \ < \ 1 \ .
\end{eqnarray}
We assumed that $p < 1/2$. Therefore by (\ref{estim-large-eps}), our conclusion is true, in fact, for
\begin{equation}\label{eps-eps-cr}
\varepsilon
\geq {1}/{p}\geq 2 = \varepsilon_{cr} \ .
\end{equation}
The same result follows in the case $p \geq 1/2$, if we interchange $p$ and $1-p$ and $\mu-1$ and
$1+\varepsilon-\mu$ in the above argument.

Next we show that for any \textit{other} $\rho \in (0,1)$, i.e. for any $\rho \neq 1 - p$,
the critical $\beta_c(\rho) < +\infty$, i.e. for these densities one always has the Bose-Einstein
condensation
at low temperatures.

To this end suppose that there is $\rho^* \in (0,1)$ such that $\rho^* \neq 1 - p$, but
$\lim_{\rho\rightarrow \rho^*} \beta_c(\rho) = + \infty$. Then the
left-hand side of (\ref{bc_bin}) converges to
\begin{equation} \label{M-func}
\lim_{\beta \to \infty}F_{p,\varepsilon}(\beta, \mu)=
M_p(\mu,\varepsilon) := \frac{p}{|\mu-\varepsilon-1|} + \frac{1-p}{|\mu-1|} \ .
\end{equation}
The number of solutions of equation (\ref{bc_bin}) in the limit
$\lim_{\rho\rightarrow \rho^*} \beta_c(\rho) = + \infty$ depends on
the value of $\varepsilon > 0$, but two singular points $\mu =1$ and
$\mu= 1 + \varepsilon$ of the function (\ref{M-func}) ensure (for
nontrivial values of the probability: $p\neq 0$ and $p\neq1$) that
there are always at least \textit{two solutions}:
$\mu_{1}(\varepsilon) < 1$ and $\mu_{2}(\varepsilon) > 1 +
\varepsilon$ of equation
\begin{equation}\label{M-eq}
M_p(\mu ,\varepsilon) = 1 \ .
\end{equation}
If $\lim_{\rho\rightarrow \rho^*} \beta_c(\rho) = + \infty$, then for these two cases the equation
(\ref{rho_bin}) implies:
\begin{eqnarray*}
&& \rho^* = \lim_{\rho\rightarrow \rho^*} G_{p,\varepsilon}(\beta_c(\rho), \mu_{1}(\varepsilon) = 0 \ ,\\
&& \rho^* = \lim_{\rho\rightarrow \rho^*} G_{p,\varepsilon}(\beta_c(\rho), \mu_{2}(\varepsilon) = 1 \ .
\end{eqnarray*}
This contradicts our assumptions on $\rho^*$ and makes impossible the hypothesis
$\lim_{\rho\rightarrow \rho^*}\beta_c(\rho)= + \infty$.

Notice that the function $M_p(\mu,\varepsilon)$ has a \textit{minimum}
$\overline{\mu}(\varepsilon)\in (1, 1 + \varepsilon)$.
If $M_p(\overline{\mu}(\varepsilon),\varepsilon) < 1 $  (which is equivalent to $\varepsilon > \varepsilon_p
:=1 + 2 \sqrt{p(1-p)}$), then equation (\ref{M-eq}) has
\textit{two} complementary  solutions $\mu_{\mp}(\varepsilon)$:
\begin{equation}\label{mu-mp}
\mu_{\mp}(\varepsilon) = \frac{\varepsilon+3}{2} - p \mp
\sqrt{\left(\frac{\varepsilon-1}{2}\right)^2 - p(1-p)} \ ,
\end{equation}
such that
\begin{equation*}
1 < \mu_{-}(\varepsilon) < \overline{\mu}(\varepsilon) < \mu_{+}(\varepsilon)< 1 + \varepsilon  \ .
\end{equation*}
If $\lim_{\rho\rightarrow \rho^*} \beta_c(\rho) = + \infty$,
then for these two solutions equation (\ref{rho_bin}) implies:
\begin{equation*}
\rho^* = \lim_{\rho\rightarrow \rho^*} G_{p,\varepsilon}(\beta_c(\rho), \mu_{\mp}(\varepsilon)) = 1 - p \ ,
\end{equation*}
This again contradicts our assumption about $\rho^*$, and thus
proves the assertion: $\beta_{c}(\rho) < + \infty$ for any $\rho
\neq 1 - p$.

Notice that by (\ref{mu-mp}) the equation $M_p(\overline{\mu}(\varepsilon),\varepsilon) = 1 $
has a unique solution
$\varepsilon = \varepsilon_{p}\leq \varepsilon_{cr} = 2$, and one obtains
$M_p(\overline{\mu}(\varepsilon),\varepsilon) > 1 $ for all $\varepsilon < \varepsilon_p$,
which excludes complementary
solutions $\mu_{\mp}(\varepsilon)$. On the other hand, if
\begin{equation}\label{crit-eps-max}
\varepsilon > \varepsilon_{cr}= \max_{p} \  \varepsilon_p = \varepsilon_{p=1/2} \ ,
\end{equation}
there are \textit{always} complementary solutions (\ref{mu-mp}). This may \textit{restrict}
the values of $\rho$, for which we have bounded critical
$\beta_c(\rho)$, to a certain domain of densities.

To this end we consider first the $\rho\,$-independent equation
(\ref{bc_bin}). Notice that $F_{p,\varepsilon}(\beta, \mu)$ is a
monotonously increasing function of $\beta$, so there is a
\textit{unique} solution $\tilde{\beta}_c(\mu)$ of equation
(\ref{bc_bin}) for a given $\mu$, \textit{if} there is one.

Since $(\tanh u)/u \leq 1$, then the left-hand side of
(\ref{bc_bin}) is \textit{less} than 1, for $\beta \leq 2$. On the
other hand, as $\beta \to \infty$, the left-hand side of
(\ref{bc_bin}) converges to  $M_p(\mu,\varepsilon)$. Since the
function (\ref{M-func}) is singular at $\mu = 1$ and $\mu = 1 +
\varepsilon$, a solution $2 < \tilde{\beta}_c(\mu) < + \infty$ for a
certain $\mu$ always exists, and the set of those $\mu$ is defined
by the condition:
\begin{equation}\label{mu-set}
S_{p,\varepsilon} := \{\mu \in \mathbb{R}^1 : \lim_{\beta \to \infty} F_{p,\varepsilon}(\beta, \mu)=
M_p(\mu,\varepsilon) \geq 1 \}
\end{equation}
By (\ref{M-func}) the set (\ref{mu-set}) for $\varepsilon > 0$ is a
\textit{compact} in $\mathbb{R}^{1}_{+}$. If there are no
complementary  solutions $\mu_{\mp}(\varepsilon)$, this compact is
\textit{connected},  but if
\begin{equation}\label{crit-eps}
\varepsilon > \varepsilon_{cr} \ .
\end{equation}
it contains two domains separated by a \textit{gap}:
\begin{equation*}
I(\varepsilon,p):= (\mu_{-}(\varepsilon)\,, \, \mu_{+}(\varepsilon)),
\end{equation*}
see (\ref{mu-mp}). The gap $I(\varepsilon,p) \subset (1,
1+\varepsilon)$. There is no solutions $\tilde{\beta}_c(\mu)$ for
$\mu \in I(\varepsilon,p)$ and for
\begin{equation*}
\mu < {(\varepsilon+1)}/{2} -
\sqrt{\left({(\varepsilon-1)}/{2}\right)^2 - \varepsilon(1-p)} \ ,
\end{equation*}  or for
\begin{equation*}
\mu > {(\varepsilon+3)}/{2} +
\sqrt{\left({(\varepsilon+1)}/{2}\right)^2 - \varepsilon(1-p)} \ .
\end{equation*}
Hence, for large $\varepsilon$ (\ref{crit-eps}) the set
$S_{p,\varepsilon}$ is a union of two (separated by the gap
$I(\varepsilon,p)$) bounded domains, which are vicinities of
singular points $\mu=1$ and $\mu = 1+\varepsilon$ .
is in fact {\bf not} the

To understand, how the gap in the chemical potential for solution
$\tilde{\beta}_c(\mu)$ modify the behaviour of ${\beta}_c(\rho)$, we
have to consider the $\rho\,$-dependent equation (\ref{rho_bin}).
Notice that from (\ref{rho_bin}) one obtains $\hat{\beta}_c(\mu,
\rho)$ as a function of two variables. Therefore,  ${\beta}_c(\rho)$
is a solution of equation:
\begin{equation}\label{eq-beta-crit}
\tilde{\beta}_c(\mu) = \hat{\beta}_c(\mu, \rho) \ ,
\end{equation}
which in fact connects $\mu$ and $\rho$: $\overline{\mu}(\rho)$,
i.e. ${\beta}_c(\rho)= \tilde{\beta}_c(\overline{\mu}(\rho)) =
\hat{\beta}_c(\overline{\mu}(\rho), \rho)$.

Clearly, the left-hand side $G_{p,\varepsilon}(\beta, \mu)$ is
increasing in $\mu$ and it tends to 0 as $\mu \to -\infty$ and to 1
as $\mu \to +\infty$. Excluding $\rho=0$ or 1, there is therefore a
\textit{unique} solution $\mu(\beta,\rho)$ of (\ref{rho_bin}) for
each value of $\beta$. As $\beta \to 0$, $G_{p,\varepsilon}(\beta,
\mu)$ tends to $1/2$ at constant $\mu$. Therefore, if $\rho \neq
1/2$
\begin{equation*}
 \lim_{\beta \to 0} \mu(\beta,\rho)  =  \pm \infty \ ,
\end{equation*}
depending on whether $\rho > 1/2 $ or $\rho < 1/2$.

On the other hand, in the limit $\beta \to \infty$, we have that
$G_{p,\varepsilon}(\beta, \mu)$: $(a)$ tends to 0, if $\mu < 1$; $(b)$
to $(1-p)/2$, if $\mu = 1$; $(c)$ to $1-p$, if $1 < \mu <
1+\varepsilon$; $(d)$ to $1 - p/2$, if $\mu = 1+\varepsilon$, and $(e)$
to 1,  if $\mu > 1+\varepsilon$.

The $(a)-(e)$ give relation between $\rho$ and $\mu$  for large $\beta$ : if
$0 < \rho < 1-p$, we must have $\mu(\beta,\rho) \to 1$ and, if $1-p
< \rho < 1$, we obtain $\mu(\beta,\rho) \to 1+\varepsilon$, for
$\beta \to \infty$. At $\rho = 1-p$, we have to use the
representation (\ref{crden}), that yields
\begin{equation}\label{mu-asympt-large-beta}
\mu(\beta, \rho = 1-p)= 1+\frac{1}{2} \varepsilon -
\frac{1}{2\beta}\ln \frac{p}{1-p} + o(\beta^{-1}) \ ,
\end{equation}
if $\beta$ is large. In particular, this justifies the remark (\ref{eps-eps-cr}) above
about $\varepsilon_{\rm cr}= 2 $, since $1 + \varepsilon/2$ lies in
the gap $I(\varepsilon,p)$ only if $\varepsilon \geq 2 = \varepsilon_{\rm cr}$, see
(\ref{mu-mp}).

Hence, it follows that for $\rho \neq 1-p \,$  two functions of $\mu$ corresponding to solutions
(\ref{eq-beta-crit}) of equations (\ref{bc_bin}), (\ref{rho_bin}) must
intersect. On the other hand, (\ref{eps-eps-cr}) proves that they can not intersect for $\rho = 1-p \,$, if
$\varepsilon > \varepsilon_{\rm cr} $. In fact, we can derive \textit{upper} bounds for $\beta_c(\rho)\, $
in the case  $\rho \neq 1-p \,$
and  $|\rho - 1+p| \,$ small.

To this end \textit{we first consider the case $\rho > 1-p$.} Let us
assume $p \leq {1}/{2}$. (The case $p > {1}/{2}$ can be studied
similarly.) Writing $\rho = 1-p+\delta/2$ we present the
equation (\ref{rho_bin}) in the form
\begin{equation}\label{rho>1-p}
p \tanh \frac{1}{2} \beta_c (\varepsilon +1 -\mu) = (1-p) \tanh
\frac{1}{2} \beta_c (\mu-1) + 2p -1 - \delta \ .
\end{equation}
Identity (\ref{rho>1-p}) implies that  $\mu > 1 + \varepsilon/2$, since otherwise we get a contradiction:
\begin{eqnarray*}
&& 1-2p + \delta = - p \tanh \frac{1}{2} \beta_c (\varepsilon +1 -\mu) + (1-p) \tanh
\frac{1}{2} \beta_c (\mu-1) \leq \\
&&- p \tanh \frac{1}{2} \beta_c (\varepsilon +1 -\mu) + (1-p) \tanh
\frac{1}{2} \beta_c (\varepsilon +1 -\mu) \leq 1-2p \ .
\end{eqnarray*}
On the other hand, for
$\varepsilon \geq 1$, one gets the upper limit $\mu < \varepsilon
+ 1$. Indeed, if we suppose the opposite: $\mu \geq \varepsilon +1 $,  then (\ref{rho_bin})
and the general fact that $\beta_c \geq 2$ (see (\ref{betag2}))
yield
\begin{eqnarray*}
1-2p+\delta &=& p \tanh \frac{1}{2} \beta_c (\mu-\varepsilon-1) +
(1-p) \tanh \frac{1}{2} \beta_c (\mu-1)
\\ && \geq (1-p) \tanh \frac{1}{2} \beta_c (\mu-1)
\geq (1-p) \tanh \varepsilon .
\end{eqnarray*}
But this is impossible for (large) $\varepsilon$ verifying:
\begin{equation}
\varepsilon > \frac{1}{2} \ln \frac{2-3p+\delta}{p-\delta} \ .
\label{lowereps}
\end{equation}
Therefore, we obtain for $\mu$ the \textit{lower} and \textit{upper} bounds:
\begin{equation}\label{mu}
1 + \varepsilon/2 < \mu < 1 + \varepsilon \ .
\end{equation}

Now identity (\ref{rho>1-p}), together with the bounds (\ref{mu}), inequality
$\tanh(u) > 1-2e^{-2u}$ and $\beta_c \geq 2$ (see (\ref{betag2})), yields the estimates:
\begin{equation}\label{estim-beta}
1 - \frac{\delta}{p} - \frac{2}{p} e^{-\varepsilon} <
\tanh \half \beta_c (\varepsilon+1-\mu) < 1- \frac{\delta}{p}.
\end{equation}
\begin{equation*}
1 > \frac{p-\delta-2
e^{-\varepsilon}}{\varepsilon + 1-\mu} + (1-p) \frac{1-2
e^{-\varepsilon}}{\mu-1} > \frac{\beta_c (p-\delta - 2
e^{-\varepsilon})}{\ln (2p/\delta)}
\end{equation*}
and hence,
\begin{equation} \label{estim-beta-above}
\beta_c < \frac{1}{p-\delta-2 e^{-\varepsilon}} \ln (2p/\delta).
\end{equation}
The upper bound (\ref{estim-beta-above}) holds for example, if $\delta < p/2$ and $\varepsilon
> \ln (4/p)$.

Now \textit{we consider the case $\rho < 1-p$ } and suppose $p \leq
{1}/{2}$, since $p > {1}/{2}$ can be studied
similarly. Then we write: $\rho = 1-p- \delta/2$. Equation
(\ref{rho_bin}) now reads as
\begin{equation}\label{rho < 1-p}
(1-p) \tanh
\frac{1}{2} \beta_c (\mu-1) = p \tanh \frac{1}{2} \beta_c
(1+\varepsilon-\mu) + 1-2 p-\delta  \ .
\end{equation}
An argument similar to the case $\rho > 1-p$ shows that
\begin{equation}\label{1-mu-1}
1 < \mu <
1+\varepsilon  \ ,
\end{equation}
if $\varepsilon$ is large enough and $\delta < 1-p$. Indeed, if we suppose the opposite:
$\mu \geq 1+\varepsilon$, then
\begin{equation*}
1-2p-\delta \geq (1-p) \tanh \half \beta_c (\mu-1)
\geq (1-p) \tanh \varepsilon  \ ,
\end{equation*}
which is impossible for
\begin{equation*}
\varepsilon > \half \ln
\frac{2-3p-\delta}{p+\delta} \ .
\end{equation*}
Similarly, if we suppose  that $\mu \leq 1$,  then (\ref{rho < 1-p}) implies
\begin{equation*}
0 > p \tanh \half \beta_c
(1+\varepsilon-\mu) + 1-2p-\delta > p \tanh \varepsilon  \  + \
(1-2p-\delta)  \ ,
\end{equation*}
which is impossible if $\delta <1-2p $ , or  if $1-2p \leq \delta < 1-p$  and
\begin{equation*}
\varepsilon > \half \ln \frac{3p-1+\delta}{1-p-\delta} \ .
\end{equation*}

Now, (\ref{rho < 1-p}) and  (\ref{1-mu-1}) imply that
\begin{equation} \label{betac-est}
\tanh \frac{1}{2} \beta_c (\mu-1) < 1 - \frac{\delta}{1-p} \ .
\end{equation}
In the case $\mu \geq 1+\half \varepsilon $ this yields immediately the upper bound :
\begin{equation}\label{upper-bound-1}
\beta_c < \frac{2}{\varepsilon} \ln
\frac{2(1-p)}{\delta} \ .
\end{equation}
On the other hand, if $1 < \mu < 1+ \varepsilon /2$, then by (\ref{rho < 1-p}) and $\beta_c \geq 2$
we obtain
\begin{eqnarray}
&& (1-p) \tanh \half \beta_c (\mu-1)> p \tanh \frac{1}{4} \beta_c
\varepsilon  + 1-2 p-\delta > \nonumber \\
&& p \tanh \frac{1}{2} \varepsilon + 1-2 p-\delta > p(1 - 2 e^{-\varepsilon}) +
1 - 2p - \delta = 1 - p - \delta - 2 p e^{-\varepsilon} \ . \label{estim-2}
\end{eqnarray}
Taking into account equation (\ref{bc_bin}) and estimates (\ref{betac-est}),
(\ref{estim-2}),  we get
\begin{equation*}
1 >  \frac{1 - p - \delta - 2 p e^{-\varepsilon}} {\mu-1}  >
\beta_c \, \frac{1 - p - \delta - 2  p e^{-\varepsilon}}{\ln(2(1-p)/{\delta})} \ ,
\end{equation*}
that gives the upper bound:
\begin{equation}\label{estim-2-2}
\beta_c < \frac{1}{1 - p - \delta - 2 p e^{-\varepsilon}}\ln
\frac{2(1-p)}{\delta} \ .
\end{equation}


\subsubsection{Bernoulli random potential for the case  $\lambda < + \infty$.}

We assume in this subsection that $\lambda > \varepsilon +1$. If the
repulsion is very large ($\lambda \gg \varepsilon +1$), the analysis
for $\rho < 1$ is then almost the same as above for $\lambda = +
\infty$, whereas for $\rho \geq 1$, which is possible only for
finite $\lambda$, one needs some more arguments.

Here we start with the  estimate the \textit{first-order} correction in $\lambda^{-1}$ to the value
of $\varepsilon_{\rm cr}(\lambda = + \infty)= 2$. With this accuracy the equations (\ref{betacrit})
and (\ref{rhocrit}) can be approximated correspondingly by
\begin{eqnarray}\label{finbin}
&& p \left( \frac{\tanh \frac{1}{2} \beta (\mu - \varepsilon-1)}{\mu
- \varepsilon-1} + \frac{1}{2 \lambda + \varepsilon + 1-\mu} \,
\frac{e^{-\beta (1+\varepsilon-\mu)/2}}{\cosh \frac{1}{2}
\beta (1+\varepsilon-\mu)} \right)  \\
&& \quad + (1-p)
\left( \frac{\tanh \frac{1}{2} \beta (\mu -1)}{\mu-1} + \frac{1}{2
\lambda + 1-\mu} \, \frac{e^{\beta (\mu-1)/2}}{\cosh
\frac{1}{2} \beta (\mu-1)} \right) = 1  \ , \nonumber
\end{eqnarray}
and by (\ref{rho_bin}) as above.

To see this, note that if $\rho < 1$, the dominant contribution in (\ref{rhocrit})
must come from the $n=1$ term, i.e. we must have $h_1 < h_2$, so
$\mu < 1+2\lambda+\varepsilon$. The other terms in (\ref{rhocrit})
are then exponentially small and can be neglected, which leads
again to (\ref{rho_bin}).

Now, because of the presence of $e^{-\beta h_1}$ in the $n=2$ term of (\ref{ptilde}),
it cannot be neglected in (\ref{betacrit}) and we obtain:
\begin{eqnarray*}
&& \frac{2p}{1+e^{-\beta (1+\varepsilon-\mu)}}
\left\{ \frac{e^{-\beta (1+\varepsilon-\mu)}-1}{\mu-1-\varepsilon}
+ 2 \frac{e^{-\beta (1+\varepsilon -
\mu)}}{1+2\lambda+\varepsilon-\mu} \right\} \nonumber
\\ && + \frac{2(1-p)}{1+e^{-\beta (1-\mu)}}
\left\{ \frac{e^{-\beta (1-\mu)}-1}{\mu-1} + 2 \frac{e^{-\beta
(1-\mu)}}{1+2\lambda-\mu} \right\} = 2  \ ,
\end{eqnarray*}
which is the same as (\ref{finbin}).

Similar to (\ref{M-func})  the gap equation for $1 < \mu < 1 + \varepsilon$ can be obtained
from (\ref{finbin})
in the limit $\beta \to \infty$:
\begin{equation}\label{eq-M-large-lambda}
\frac{p}{\varepsilon+1-\mu} + (1-p) \left(
\frac{1}{\mu-1} + \frac{2}{2 \lambda+1-\mu} \right) = 1.
\end{equation}
If $\rho = 1-p$, then by (\ref{rho_bin}) and (\ref{mu-asympt-large-beta}) we again obtain the limit:
$\mu \to 1 + \frac{1}{2} \varepsilon$ for $\beta \to \infty$. Inserting this limit
into (\ref{eq-M-large-lambda}) we obtain
\begin{equation}\label{rho=1-p}
\frac{2}{\varepsilon} + \frac{2(1-p)}{2
\lambda -\frac{1}{2} \varepsilon} = 1 \ .
\end{equation}
Hence, by the reasoning similar to those after (\ref{mu-asympt-large-beta}), we obtain the
critical value of the Bernoulli random potential $\varepsilon_{\rm cr}(\lambda)$ the expression:
\begin{equation}\label{eps-crit-corr}
\varepsilon_{\rm cr}(\lambda) \approx
\frac{2}{1 - (1-p)/ \lambda} = 2 + 2(1-p)/\lambda + \ldots \ ,
\end{equation}
which takes into account that $\lambda$ is large but \textit{finite}.

Another observation, which is related to the finiteness of $\lambda$, concerns the value $\beta_{c}(\rho=1)$.
For hard-core bosons the arguments in the Sect.4.2.1 show that this value is \textit{infinite}
and the corresponding
values of the chemical potential must be greater than $1+\varepsilon$, see (\ref{rhocrit}).
Now for finite $\lambda$ and $\mu > 1+\varepsilon$ the limit of (\ref{finbin}), when $\beta \to
\infty$, reads as:
\begin{equation}\label{newbetac}
p \left( \frac{1}{\mu-\varepsilon-1} +
\frac{2}{2\lambda+1+\varepsilon-\mu} \right) + (1-p) \left(
\frac{1}{\mu-1} + \frac{2}{2 \lambda+1-\mu} \right) = 1.
\end{equation}
If  $\rho \geq 1 $, then  we need to reconsider the density
equation (\ref{rhocrit}), which has the form:
\begin{equation}
\rho = p \
\frac{\sum_{n=1}^\infty n \ e^{-\beta
h_n(\mu-\varepsilon,\lambda)}}{\sum_{n=0}^\infty e^{-\beta
h_n(\mu-\varepsilon,\lambda)}} + (1-p)\
\frac{\sum_{n=1}^\infty n \ e^{-\beta
h_n(\mu,\lambda)}}{\sum_{n=0}^\infty e^{-\beta
h_n(\mu,\lambda)}}. \label{newrhocrit}
\end{equation}
Notice that if $\beta \to +\infty$, then by  (\ref{h})  and (\ref{newrhocrit}) one obtains the
following limits:
$\rho \to 1$, when $\mu \in (1+\varepsilon, 1+2\lambda)$ , $\rho \to 2-p$, when $\mu \in (1+2\lambda,
1+2\lambda+\varepsilon)$, and $\rho \to 2$, when $\mu \in
(1+2\lambda+\varepsilon,1+4\lambda)$.

Therefore, at $\rho = 1$ for large $\beta$ we can ignore in (\ref{newrhocrit}) the terms higher than $h_2$,
see (\ref{h}), and write in this limit:
\begin{eqnarray}
1 &\approx& p \left\{ \frac{e^{-\beta(1+\varepsilon-\mu)} + 2
e^{-2\beta(1+\lambda+\varepsilon-\mu)}}{1 + e^{-\beta
(1+\varepsilon-\mu)} + e^{-2 \beta(1 + \lambda +
\varepsilon-\mu)}} \right\} \nonumber
\\ && \qquad + (1-p) \left\{
\frac{e^{-\beta(1-\mu)} + 2 e^{-2\beta(1+\lambda-\mu)}}{1 +
e^{-\beta (1-\mu)} + e^{-2\beta(1 + \lambda-\mu)}} \right\} \nonumber
\\ &=& p \left\{ \frac{1 + 2 e^{-\beta (1+\
2\lambda+\varepsilon-\mu)}}{1 + e^{-\beta(\mu-1-\varepsilon)} +
e^{-\beta(1+2\lambda+\varepsilon- \mu)}} \right\} \nonumber
\\ && \qquad
+ (1-p) \left\{ \frac{1 +
2e^{-\beta(1+2\lambda-\mu)}}{1+e^{-\beta(\mu-1)} +
e^{-\beta(1+2\lambda-\mu)}} \right\} \label{pho=1-eq}
\\ &\approx& 1 + p
\left( e^{-\beta(1+2\lambda + \varepsilon-\mu)} -
e^{-\beta(\mu-1-\varepsilon)} \right) \nonumber
\\ && \qquad + (1-p)
\left( e^{-\beta(1+2 \lambda-\mu)} - e^{-\beta(\mu-1)} \right).\nonumber
\end{eqnarray}
This yields
\begin{equation*}
e^{2 \beta \mu}
\approx e^{2 \beta (1+\lambda)} \frac{1-p+pe^{\beta
\varepsilon}}{1-p+p e^{-\beta \varepsilon}} \approx \frac{p}{1-p}
e^{2 \beta (1+\lambda+\half \varepsilon)}.
\end{equation*}
The chemical potential defined by equation (\ref{newrhocrit}) therefore tends (for $\rho = 1$)
to $1 + \lambda + \frac{1}{2} \varepsilon$  as $\beta \to +\infty$.

Therefore, inserting this into (\ref{newbetac}) we obtain the
estimate for the value of \textit{repulsion} $\lambda_{c,1}$ that
ensures that $\beta_c (\rho = 1)= +\infty$  in the presence of the
random Bernoulli potential:
\begin{equation}\label{lambda-cr}
\lambda_{c,1}(\varepsilon) = \frac{1}{2} \left[ 3 + \sqrt{9 + 2 \varepsilon
(1-2p+\frac{1}{2} \varepsilon)} \right].
\end{equation}

\begin{remark}\label{rem-lambda-cr}
In the absence of disorder, i.e. if $\varepsilon = 0$, the
critical value of lambda is $\lambda_{c,1} = 3$ as opposed to
$\lambda_1 = \frac{1}{2} (3+\sqrt{8})$ as suggested in \cite{BD}.
The reason is the same as above for $\varepsilon_{\rm cr}$,
namely, the graph of $\mu(\beta,\rho)$ at $\rho=1$ tends to
$1+\lambda$ as $\beta \to +\infty$ and this lies in the gap only
if $\lambda \geq 3$. Similarly, the next critical values are given
by
\begin{equation} \lambda_{c,k}(\varepsilon=0) = 2k+1.
\end{equation}
\end{remark}
\begin{remark}\label{rho=1-p-bis}
In Sect.4.2.1 we notice a new phenomenon specific for the random
case: divergence of $\beta_c$ at $\rho = 1-p$ for hard-core
bosons, cf. Figure 1 for $p=1/2$. Instead of fixing $\lambda$,
fixing  $\varepsilon > 2$ it follows from (\ref{rho=1-p}) that
there is a critical value of the \textit{repulsion}
$\lambda_{c,1-p}(\varepsilon)$ (instead of $\varepsilon$ as in
(\ref{eps-crit-corr})) so that $\beta_c (\rho = 1 - p)$ diverges
for $\lambda \geq \lambda_{c,1-p}(\varepsilon)$ in the presence of
the random Bernoulli potential:
\begin{equation}\label{lambda-cr-1-p}
\lambda_{c,1-p}(\varepsilon) = \frac{\varepsilon}{4} +
\frac{\varepsilon(1-p)}{\varepsilon-2}.
\end{equation}
This critical value is not evident from Figure 1 as $\varepsilon =
2$.
\end{remark}
\begin{remark}\label{rem-beta-cr}
In Sect.4.1 we remarked that the critical temperature for
\textit{free} bosons increases due to disorder. We also remarked
that for the interacting case this is a more subtle matter, since
it depends on the value of repulsion. For \textit{large}
repulsions close to e.g. $\lambda_{c,1}(\varepsilon = 0) = 3$, we
get by (\ref{lambda-cr}) that
\begin{equation}\label{temp-rand-incr}
\beta_c (\rho = 1; \lambda = 3, \varepsilon > 0) < \beta_c (\rho = 1; \lambda = 3, \varepsilon = 0) =
+ \infty \ .
\end{equation}
This \textit{lowering} of $\beta_c (\rho = 1)$ can be explained
intuitively as follows. At density $\rho = 1$, there is one
particle per site. If $\varepsilon = 0$ there is a penalty for a
particle to jump to an already occupied site, so the preferred
state is where the particles are at fixed sites, which is almost
an eigenstate of the number operators $n_x$ for each site. This
prevents Bose condensation. (This argument was presented also in
\cite{BD}.) However, if $\varepsilon > 0$, then the lattice splits
into two parts with energies 0 and $\varepsilon$, and a particle
jumping from a site with energy $\varepsilon$ to a site with
energy 0 loses an energy $\varepsilon$, which counteracts the gain
of $\lambda$. This creates more freedom of movement and therefore
promotes Bose condensation. On the other hand, for a fractional
value of the $\rho$ in the neighbourhood of $\rho = 1-p$, the
critical temperature decreases with increasing $\varepsilon$ as
can be seen from Figure 1.
\end{remark}

Now consider the case $\rho > 1$. From equation
(\ref{newrhocrit}) we see that at fixed $\rho \in (1,2-p)$, $\mu
\to 1+2\lambda$ and for $\rho \in (2-p,2)$, $ \mu \to 1+2
\lambda+\varepsilon$ as $\beta \to \infty$.

For the case $\rho = 2-p$, we have to expand (\ref{newrhocrit}), as above for $\rho=1$,
see (\ref{pho=1-eq}), but to take into account that $ \mu \in (1+2\lambda, 1+2\lambda+\varepsilon)$:
\begin{eqnarray}
\rho &\approx & p \left\{ \frac{1 + 2 e^{-\beta (1+\
2\lambda+\varepsilon-\mu)}}{1 + e^{-\beta(\mu-1-\varepsilon)} +
e^{-\beta(1+2\lambda+\varepsilon- \mu)}} \right\} \label{pho>1-eq}
\\ && \qquad
+ (1-p) \left\{ \frac{e^{\beta(1+2\lambda-\mu)} +
2}{1+e^{\beta(1 +2\lambda -\mu)} +
e^{-2\beta(\mu - 1 -\lambda)}} \right\}  \nonumber \\
&\approx& 2-p + p \left(e^{-\beta(1+\varepsilon+2 \lambda - \mu)}
- e^{-\beta (\mu-1-\varepsilon)} \right) - \nonumber \\
&& (1-p) e^{-\beta (\mu -1-2 \lambda)} - 2(1-p) e^{-2\beta (\mu - 1 - \lambda)}. \nonumber
\end{eqnarray}
This yields that $e^{-\beta(\mu-1-2 \lambda)} \approx
e^{-\beta(1+\varepsilon+2 \lambda-\mu)}p/(1-p)$ for large $\beta$, i.e.
$\mu \to 1+2 \lambda + \frac{1}{2} \varepsilon$, if $\rho = 2-p$ and $\beta \to \infty$.

For $\mu \approx 1+2\lambda+\half \varepsilon$, one has
$h_1(\mu-\varepsilon, \lambda) < h_2(\mu-\varepsilon, \lambda)$. So
that the $p$-terms in (\ref{finbin}) are unchanged, but $h_1(\mu,
\lambda) > h_2(\mu, \lambda) < h_3(\mu, \lambda)$, if $\lambda >
\varepsilon /4$, which corresponds to our initial hypothesis about
the value of repulsion: $\lambda > 1 + \varepsilon$. Hence, the
$(1-p)$-terms are now dominated for large $\beta$ by $n=2$ and
(\ref{finbin}) read as
\begin{eqnarray*}
&& \frac{p}{1+e^{-\beta (1+\varepsilon-\mu)}} \left\{
\frac{e^{-\beta (1+\varepsilon-\mu)}-1}{\mu-1-\varepsilon} + 2 \
\frac{e^{-\beta (1+\varepsilon - \mu)}}{1+2\lambda+\varepsilon-\mu}
\right\} \nonumber
\\ && + \frac{1-p}{e^{-\beta (1-\mu)} + e^{-2\beta(1-\mu+\lambda)}}
\left\{ 2 \ \frac{e^{-2\beta(1-\mu+\lambda)} - e^{-\beta
(1-\mu)}}{\mu-1-2\lambda} + 3 \
\frac{e^{-2\beta(1-\mu+\lambda)}}{1+4\lambda-\mu} \right\} \approx
1,
\end{eqnarray*}
In the limit $\beta \to \infty$ we obtain from this relation the gap
equation
\begin{eqnarray}\label{gap-rho>1}
&&p \left( \frac{1}{\mu-1-\varepsilon} +
\frac{2}{1+\varepsilon+ 2\lambda-\mu} \right) + \\
&&(1-p) \left(\frac{2}{\mu-1-2\lambda} + \frac{3}{1+4\lambda-\mu} \right) = 1 \nonumber \ .
\end{eqnarray}
Inserting $\mu=1+2\lambda+\half \varepsilon$ into (\ref{gap-rho>1})
leads to
\begin{equation} \label{eq-eps-cr-2}
\half \varepsilon^2 - (2 \lambda-1+2p)
\varepsilon + 8\lambda  = 0.
\end{equation}
Solutions of (\ref{eq-eps-cr-2}) are:
\begin{equation} \label{eps-cr-1-2}
{\varepsilon_{\rm{cr}, \pm}}^{(2)}= (2 \lambda-1+2p) \pm \sqrt{(2\lambda-1+2p)^2 - 16 \lambda} \ .
\end{equation}
Hence, there is a solution that for large $\lambda$ has the form:
\begin{equation}\label{eps-cr-(2)}
\varepsilon_{\rm cr}^{(2)}(\lambda) = 4 \left(1 + \frac{1-2p}{2\lambda}\right) + \ldots  \ ,
\end{equation}
or other way around, for a given $\varepsilon$ we have:
\begin{equation}\label{lambda-cr-(2)}
\lambda_{c,\rho=2-p}(\varepsilon)= \frac{ 2(2p-1)}{( \varepsilon - 4)} \ .
\end{equation}
Clearly, this critical value only applies if $\varepsilon > 4$ and
$p > 1/2$. The top graph of Figure 1 illustrates this behaviour at
$\rho =1.5$ for $\varepsilon = 4.5$ and $\lambda = 10$.
\vskip0.5cm

The critical $\beta_c(\rho)$ for the Bernoulli distribution with
$p={1}/{2}$ and $\varepsilon =2$ is shown in Figure 1 for a number
of values of $\lambda$. Notice in particular that $\varepsilon <
\varepsilon_{cr}(\lambda)$, see (\ref{eps-crit-corr}),  for all
finite $\lambda$, so that  $\beta_c (\rho = {1}/{2}) < +\infty$.

Also, for $\lambda= 3.3 $, one obtains $\beta_c(\rho = 1) < +\infty
$ because $ 3.3 < \lambda_{c,1}(\varepsilon =2) = (3+\sqrt{13})/2 $,
see (\ref{lambda-cr}).\\

\begin{figure}[h]
\centerline{\includegraphics[angle=0,width=9.5cm]{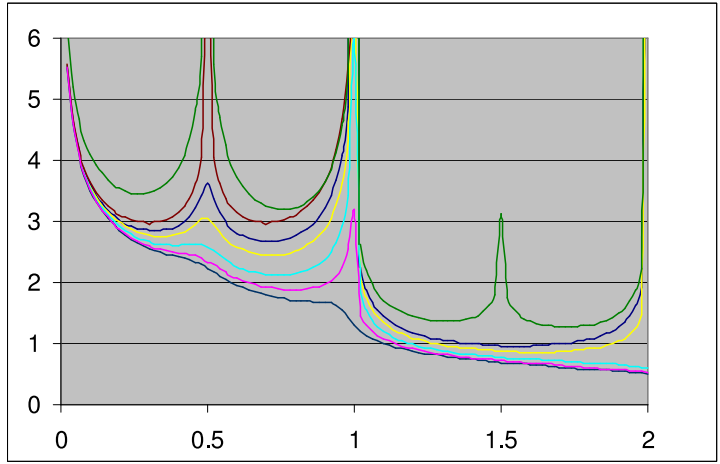}}
\centerline{\parbox{11cm}{\caption{\emph{$\beta_c$ as a function
of the density $\rho$ in the case of averaging over two energies:
0 and  $\varepsilon =2$  with equal probabilities, for various
values of $\lambda$: $\lambda = 3, 3.3, 4, 6, 10$ and $+\infty$.
The top graph corresponds to the case $\varepsilon = 4.5$ and
$\lambda = 10$. }}}} \label{fig1}
\end{figure}

\subsubsection{Trinomial distribution: $\lambda = + \infty$.}

We also briefly consider the trinomial distribution, taking for
simplicity equal probabilities, i.e.
\begin{equation}\label{3-eps}
\varepsilon^\omega = \left\{
\begin{array}{l@{ \ \ \ \mbox{Pr =}}l} 0 &  \ {1}/{3}  \\
\frac{1}{2} \varepsilon & \ {1}/{3}  \\ \varepsilon & \ {1}/{3} \ \ .
\end{array} \right.
\end{equation}
For hard-core bosons, $\lambda = +\infty$, equation (\ref{betacrit1}) for the critical value of
$\beta_c (\rho)$ takes the form:
\begin{equation}\label{3-mu}
\frac{1}{3} \left[ \frac{\tanh
\frac{1}{2} \beta(\mu-1)}{\mu-1} + \frac{\tanh \frac{1}{2} \beta
(\mu-1-\frac{1}{2} \varepsilon)}{\mu-1-\frac{1}{2} \varepsilon} +
\frac{\tanh \frac{1}{2} \beta
(\mu-1-\varepsilon)}{\mu-1-\varepsilon} \right] = 1 \ .
\end{equation}
The density equation (\ref{rhocrit1}) now reads as
\begin{equation}\label{3-pho}
\rho =
\frac{1}{2} + \frac{1}{6} \left( \tanh \frac{1}{2} \beta(\mu-1) +
\tanh \frac{1}{2} \beta (\mu-1-\frac{1}{2} \varepsilon) + \tanh
\frac{1}{2} \beta (\mu-1-\varepsilon) \right).
\end{equation}
Then by the same analysis as in Sect.4.2.1 one gets from (\ref{3-pho}):
\begin{equation*}
\lim_{\beta \to \infty} \rho(\beta,\mu) = \left\{
\begin{array}{l@{\mbox{ \ \ if }}l} 0 &  \ \ \mu < 1 \\ {1}/{6} &  \ \ \mu =
1 \\ {1}/{3} &  \ \ 1 < \mu < 1+ \varepsilon /2 \\
{1}/{2} &  \ \ \mu =
1 + \varepsilon /2 \\ {2}/{3} &  \ \ 1+  \varepsilon /2 < \mu < 1+\varepsilon \\
{5}/{6} &  \ \ \mu = 1+\varepsilon \\ 1 &  \ \ \mu > 1+\varepsilon \ \ .
\end{array} \right.
\end{equation*}
Other way around this can be also expressed as:
\begin{equation*}
\lim_{\beta \to
\infty} \mu(\beta,\rho) = \left\{
\begin{array}{l@{\mbox{  \ \ if }}l} 1 &  \ \ 0 < \rho < {1}/{3};
\\ 1+  \varepsilon /4 &  \ \ \rho = {1}/{3}; \\ 1 +
\varepsilon /2 &  \ \ {1}/{3} < \rho < {2}/{3}; \\ 1+ {3}\varepsilon /4
&  \ \ \rho = {2}/{3}; \\ 1+\varepsilon &  \ \ \rho >
{2}/{3}.
\end{array} \right.
\end{equation*}
Again, similar to the reasoning in Sect.4.2.1, the inserting of $\mu = 1+ \varepsilon /4$
or $\mu = 1+ {3}\varepsilon/{4}$ into the limiting equation (\ref{3-mu}) for
$\beta \rightarrow +\infty$ yields the critical value of the random potential:
\begin{equation}
\varepsilon_{\rm cr} = \frac{28}{9} \ .
\end{equation}
Therefore, (similar to the Bernoulli case for $\rho = 1/2$)  the condensation of hard-core bosons
is absent at densities $\rho = 1/3$ and $\rho = 2/3$,
if $\varepsilon \geq \varepsilon_{\rm cr}$.
This phenomenon of course persists for $\lambda < +\infty$ and there are
similar suppressions of Bose condensation at $\rho = {4}/{3},
{5}/{3},$ etc.,  if $\varepsilon$ is large enough.

\subsubsection{Trinomial distribution: $\lambda < + \infty$.}

For $\lambda < +\infty$ there is a similar enhancement of Bose
condensation at $\rho = 1$ as for the Bernoulli distribution, but the
effect is stronger. This can be seen in Figure 2. The explanation is
similar to that in Remark \ref{rem-beta-cr}, except now the lattice splits into 3 equal
parts with energies 0,
$\varepsilon /2$ and $\varepsilon$. Particles can jump from
a singly-occupied site with energy $\varepsilon$ to a
singly-occupied site with energy 0 or $\varepsilon/2$, thus
compensating for the energy penalty of $\lambda$ due to double
occupation.

\begin{figure}[h]
\centerline{\includegraphics[angle=0,width=9.5cm]{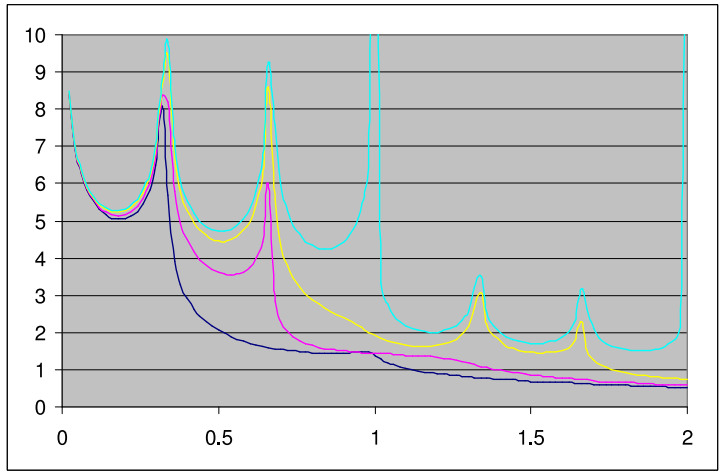}}
\centerline{\parbox{11cm}{\caption{\emph{$\beta_c$ as a function of
the density $\rho$ in the case of a trinomial distribution with
width $\varepsilon = 10$ for $\lambda = 3, 4, 6$ and $8$. }}}}
\label{fig2}
\end{figure}

By equation (\ref{rhocrit1}) for (\ref{3-eps}) we obtain that at $\rho =1$, $\mu(\beta,\rho) \to
1+\lambda+ \varepsilon/2$ as $\beta \to + \infty$. The gap
equation (\ref{3-mu}) then reduces to
\begin{equation*}
\frac{1}{\lambda-\varepsilon/2} + \frac{1}{\lambda} +
\frac{1}{\lambda+ \varepsilon/2} = 1 \ .
\end{equation*}
We can solve it for $\varepsilon$ provided $\lambda \geq 3$:
\begin{equation}
 \varepsilon_{\rm cr}(\lambda) = 2 \lambda
\sqrt{\frac{\lambda-3}{\lambda-1}} \ .
\end{equation}
Thus, Bose condensation is absent, if $\lambda \geq 3$ and
$\varepsilon \leq \varepsilon_{\rm cr}(\lambda)$.

Figure 2 shows $\beta_c(\rho)$ for a
fixed $\varepsilon=10$ and for values of $\lambda \geq 3$. Then
$\varepsilon \geq \varepsilon_{\rm cr}(\lambda = 3,4,6)$, but
$\varepsilon < \varepsilon_{\rm cr}(\lambda = 8)= 13.52$, which excludes condensation at $\rho =1$
in the latter case.

\subsubsection{General discrete distribution.}

The same phenomena persist for higher numbers of random potential
energy values, but the critical value $\varepsilon_{\rm
cr}(\lambda)$ becomes rapidly very large. Figure 3 shows the case of
a distribution with equals probabilities Pr $=1/10$ at 10
equidistant values of $\varepsilon^\omega$ (with maximal value
$\varepsilon=10$) for $\lambda = 8$. Clearly, condensation is
suppressed at $\rho = {1}/{10}, \dots, {9}/{10}$ and $\rho =1,2$ but
not at corresponding fractional values above 1, cf. Figure 2.

\begin{figure}[h]
\centerline{\includegraphics[angle=0,width=9.5cm]{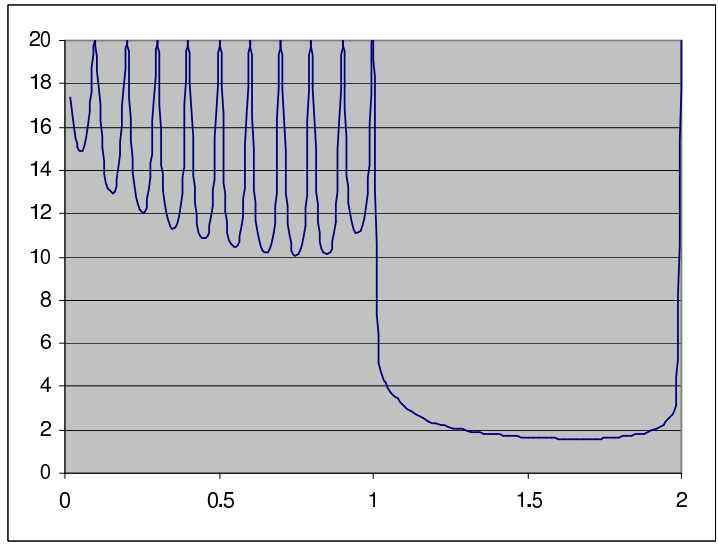}}
\centerline{\parbox{11cm}{\caption{\emph{$\beta_c$ as a function of
the density $\rho$ in the case of averaging over 10 energy values
with width $\varepsilon = 10$ for $\lambda = 8$. }}}} \label{fig3}
\end{figure}

\subsection{Continuous distribution}

\subsubsection{The case $\lambda = +\infty$.}

Consider a random potential with  \textit{homogeneous} distribution between 0 and $\varepsilon$. In
case $\lambda = +\infty$ the equations (\ref{betacrit1}) and
(\ref{rhocrit1}) become
\begin{equation}\label{cont-mu-eq}
\frac{1}{\varepsilon} \int_0^\varepsilon \frac{\tanh \frac{1}{2}
\beta (\mu-1-x)}{\mu-1-x} dx = 1
\end{equation}
and
\begin{equation}\label{cont-rho-eq}
 \frac{1}{\varepsilon} \int_0^\varepsilon \tanh \frac{1}{2} \beta
(\mu-1-x) dx = 2\rho -1 \ .
\end{equation}
The latter has sense only for $0 \leq \rho \leq 1$ and can be solved
exactly for $\mu$:
\begin{equation*}
\frac{2}{\beta \varepsilon}
\ln \frac{e^{\beta (\mu-1)/2} + e^{-
\beta (\mu-1)/2}} {e^{ \beta (\mu-1-\varepsilon)/2} +
e^{- \beta (\mu-1-\varepsilon)/2}} = 2 \rho-1 \ ,
\end{equation*}
and hence
\begin{equation}\label{cont-mu-expr}
\mu(\beta, \rho) = 1+\frac{1}{2} \varepsilon + \frac{1}{\beta} \ln
\frac{\sinh \frac{1}{2} \beta \rho \varepsilon}{\sinh \frac{1}{2}
\beta (1-\rho)\varepsilon} \ .
\end{equation}
As $\beta \to +\infty$, the expression (\ref{cont-mu-expr}) takes
the form
\begin{equation}\label{cont-mu-expr-lim}
\lim_{\beta \to +\infty}\mu(\beta, \rho) :=\overline{\mu}(\rho) =
1 + \varepsilon \rho \ , \ \ \  0 < \rho < 1 \ ,
\end{equation}
whereas $\overline{\mu}(\rho = 0) \in (- \infty, 1]$ and
$\overline{\mu}(\rho = 1) \in [1 + \varepsilon, + \infty)$ for extreme values of density, i.e.,
the inverse function is
\begin{equation}\label{cont-pho-lim}
\overline{\rho}(\mu) = \left\{
\begin{array}{l@{ \ \ \ }l}  0 &  \ \mu \leq 1  \\
(\mu -1)/\varepsilon  & \ 1< \mu < 1 + \varepsilon  \\ 1 & \ 1 + \varepsilon \leq \mu \ \ .
\end{array} \right.
\end{equation}
Then by (\ref{cont-mu-eq}) and (\ref{cont-pho-lim}) we obtain for $\rho = 1$ in the limit
$\beta \to +\infty$:
\begin{equation*}
1 =\frac{1}{\varepsilon} \int_0^\varepsilon \frac{1}{\mu-1-x} dx \ ,
\end{equation*}
or we get explicitly the value of the chemical potential
\begin{equation*}
\overline{\mu}(\rho = 1) = 1 + \frac{\varepsilon}{1 - e^{-\varepsilon}}> 1 + \varepsilon \ ,
\end{equation*}
and similarly
\begin{equation*}
\overline{\mu}(\rho = 0) = 1 - \frac{\varepsilon e^{-\varepsilon}}{1 - e^{-\varepsilon}} < 1  \ .
\end{equation*}
Hence, for hard-core bosons  the critical $\beta_c(\rho)$ is
infinite at extreme densities $\rho = 0, 1$ for \textit{any value} $\varepsilon > 0$ of the
uniform continuous distribution.

If $0<\rho<1$, then solution of the equation (\ref{cont-rho-eq}) in the limit $\beta \to +\infty$ is
(\ref{cont-mu-expr-lim}), whereas the integral in (\ref{cont-mu-eq}) diverges. Therefore, if the
critical $\beta_c(0<\rho<1)$ exist, it must be bounded. Moreover, since $(\tanh u)/{u} \leq 1$,
by (\ref{cont-mu-eq})we get for it a bound from below: $2 < \beta_c(0<\rho<1)$.

To prove the existence and uniqueness of $\beta_c(0<\rho<1)$
consider first (\ref{cont-rho-eq}) for $\rho \leq \frac{1}{2}$.
Then by virtue of (\ref{cont-mu-expr}) for \textit{any} finite
$\beta$ the solution $\mu(\beta, \rho)$ increases  from $-\infty$
to $1 + \varepsilon/2$ when $\rho$ changes from $0$ to $1/2$. For
this variation of chemical potential the integral in the left-hand
side of (\ref{cont-mu-eq}) increases monotonously from $0$ to its
\textit{maximal} value given by
\begin{equation}\label{int-max}
I(\beta, \mu = 1 + \varepsilon/2) = \frac{1}{\varepsilon}
\int_0^\varepsilon \frac{\tanh \frac{1}{2} \beta (x -
\varepsilon/2 )}{x - \varepsilon/2} dx \ .
\end{equation}
Indeed,
\begin{equation*}
\partial_{\mu}I(\beta, \mu) = \frac{1}{\varepsilon} \left( \frac{\tanh \frac{1}{2} \beta
(\mu-1)}{\mu-1} - \frac{\tanh \frac{1}{2} \beta
(\mu-1-\varepsilon)}{\mu-1-\varepsilon} \right) \geq 0
\end{equation*}
for $\mu \leq 1 + \varepsilon/2$. The integral in (\ref{int-max})
is obviously an increasing function of $\beta$. So, there exist
$\beta_0 > 2$ such that the maximal value of integral $I(\beta_0,
\mu = 1 + \varepsilon/2) \geq 1$. Hence, for any $\beta \geq
\beta_0 $ there is a \textit{unique} density $0 <
\overline{\rho}(\beta) \leq 1/2$ such that
\begin{equation}\label{solut-pho}
I(\beta, \mu(\beta, \overline{\rho}(\beta)) = 1 \ .
\end{equation}
Notice that by (\ref{cont-mu-expr}) $\mu(\beta, \rho)$ is
increasing of the both arguments: $\beta$ and $0<\rho\leq 1/2$.
Hence, to satisfy (\ref{solut-pho}) $\overline{\rho}(\beta)$ must
be decreasing function of $\beta$, i.e., the \textit{inverse}
function $\beta_c= \beta_c (\rho)$ is also a decreasing with
$\lim_{\rho \rightarrow 0}\beta_c (\rho)= + \infty$ and
$\lim_{\rho \rightarrow 1/2}\beta_c (\rho)\geq \beta_0$.

Similar arguments are valid for $1/2 \leq \rho < 1$. Whereas
$\mu(\beta, \rho)$ is still increasing function of $\rho$, the
integral $I(\beta, \mu)$ now decreases with $\mu$ from its maximal
value (\ref{int-max}) to $0$. Therefore, $\beta_c= \beta_c (\rho)$
is a monotonously increasing function of $\rho$ with $\lim_{\rho
\rightarrow 1/2}\beta_c (\rho)\geq \beta_0$ and $\lim_{\rho
\rightarrow 1}\beta_c (\rho)= + \infty$ , i.e. with a minimum at
$\rho =1/2$ as we have seen for discrete distributions and
hard-core bosons.

\subsubsection{The case of large $\lambda < +\infty$.}

By virtue of equations (\ref{betacrit}) and (\ref{rhocrit}),
for $\lambda < +\infty$, the Bose condensate is still suppressed
at $\rho = k$.

The analysis is very similar to the case
$\varepsilon = 0$. In the limit $\beta \to +\infty$ by (\ref{rhocrit}) the density
tends to ($k=0,1,\dots$)
\begin{equation*}
\rho(\mu,\beta) \to \left\{
\begin{array}{l@{\mbox{ if }}l} 0 & \mu < 1 \\ k +
\frac{1}{\varepsilon}(\mu-1-2k\lambda) & 1 + 2k\lambda < \mu < 1 +
2k\lambda + \varepsilon \\ k+1 & 1 + 2k\lambda+\varepsilon < \mu <
1 + 2(k+1) \lambda. \end{array} \right.
\end{equation*}
(To see this note that if
$1+2k\lambda < \mu < 1+2k\lambda+\varepsilon$ then the term
$e^{-\beta h_{k+1}}$ dominates for $x < \mu-1-2k\lambda$ and the
term $e^{-\beta h_k}$ dominates for $x > \mu-1-2k\lambda$.)
Clearly, if $0 < \rho < 1$ then for solution of (\ref{rhocrit}) one gets
as above: $\mu(\beta,\rho) \to 1 + \rho \varepsilon$ when
$\beta \to +\infty$. If $\rho = 1$, we need to approximate (\ref{rhocrit}) more
carefully:
\begin{eqnarray*}
1 &\approx& \frac{1}{\varepsilon} \int_0^\varepsilon \frac{
e^{\beta (\mu-1-x)} + 2 e^{2\beta (\mu-1-x-\lambda)}}{1 + e^{\beta
(\mu-1-x)} + e^{2 \beta (\mu-1-x-\lambda)}} dx \nonumber \\ &\approx&
\frac{1}{\varepsilon} \int_0^\varepsilon \left[ 1 + e^{-\beta
(1+x+2 \lambda -\mu)} - e^{-\beta (\mu-1-x)} \right] dx.
\end{eqnarray*}
Working out the integral, we find that $\mu(\beta,\rho=1) \to 1 +
\lambda + \frac{1}{2} \varepsilon$ as $\beta \to +\infty$. More generally, if $\rho = k$,
$\mu(\beta,\rho=k) \to 1 + (2k-1) \lambda + \frac{1}{2} \varepsilon$. For large
$\beta$, the gap equation  (\ref{betacrit}) becomes
\begin{equation*}
\frac{1}{\varepsilon} \int_0^\varepsilon
\left\{ \frac{k}{\mu - 1-2(k-1)\lambda) - x} +
\frac{k+1}{1+2k\lambda+x+2\lambda-\mu} \right\} dx = 1.
\end{equation*}
Inserting $\mu = 1+(2k-1) \lambda + \frac{1}{2}\varepsilon$ we obtain that
\begin{equation*}
\frac{1}{\varepsilon} \int_0^\varepsilon
\left\{ \frac{k}{\lambda + \frac{1}{2} \varepsilon -x} +
\frac{k+1}{\lambda - \frac{1}{2} \varepsilon + x} \right\} dx = 1.
\end{equation*}
This gives for the critical values of repulsion:
\begin{equation}\label{crit-lambda-k}
\lambda_{c,k}(\varepsilon) = \frac{1}{2} \varepsilon \,
\frac{e^{\varepsilon/(2k+1)} + 1}{e^{\varepsilon/(2k+1)}-1}.
\end{equation}
It is easy to see that this is larger than for non-random case
$\lambda_{c,k}(0) = 2k+1$ and agrees with the value mentioned
above at $\varepsilon = 0$, see Sect.4.2.2 .

Figure~3 shows the phase diagram for $\lambda = 10$ with
$\varepsilon = 3$, taking an average over a uniform distribution corresponding to
10 equidistant random values of $\varepsilon^\omega$ in the interval $[0, 3]$. It shows that
this already approximates the continuous case quite well.

\begin{figure}[h]
\centerline{\includegraphics[angle=0,width=9.5cm]{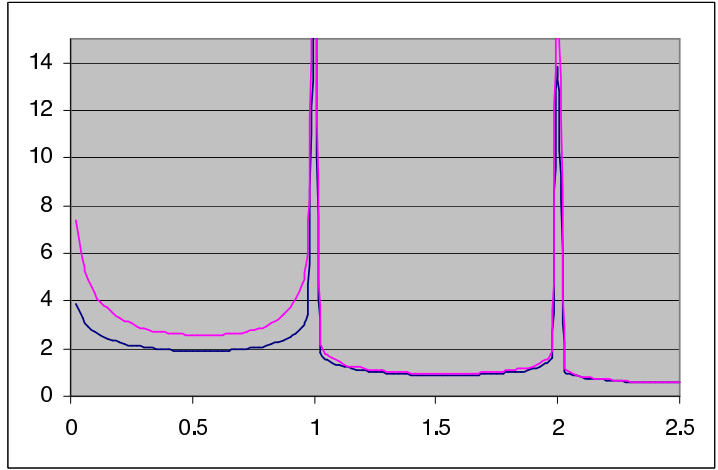}}
\centerline{\parbox{11cm}{\caption{\emph{$\beta_c$ as a function
of the density $\rho$ in the case of a near-continuous
distribution: averaging over 10 energy values with width
$\varepsilon = 3$ for $\lambda = 10$. The lower graph is the case
without randomness. }}}} \label{fig4}
\end{figure}

\subsubsection{The case of small $\lambda > 0$.}

We finally consider the case of small $\lambda$. Figure~4 shows
that, in contradistinction to the case $\lambda = 0$, for small $\lambda$ the critical
$\beta_c(\lambda, \varepsilon) > \beta_c(\lambda=0, \varepsilon=0)$, i.e. it
is \textit{larger} than that at $\varepsilon = 0$!

\begin{figure}[h]
\centerline{\includegraphics[angle=0,width=9.5cm]{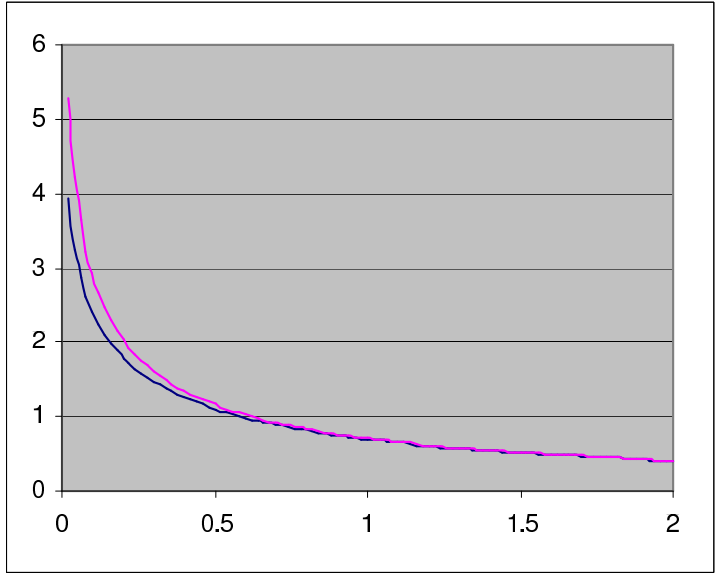}}
\centerline{\parbox{11cm}{\caption{\emph{$\beta_c$ as a function
of the density $\rho$ in the case of averaging over two energies
and width $\varepsilon = 2$ for small $\lambda = 0.1$. For
comparison, the lower graph shows the case without randomness.}}}}
\label{fig5}
\end{figure}

This can be understood as follows. Whereas in the free case $\lambda = 0$, we
must have $\mu < 0$, when $\lambda > 0$, this is no longer so. In
the limit $\lambda \to 0$, we can replace $e^{-\beta
h_n(\mu,\lambda)}$ in the expression (\ref{ptilde}) for ${\tilde
p}''(\beta, \mu, \lambda; 0)$ occurring in the gap equation (\ref{betacrit}) by
$e^{\beta (\mu-1)}$. Replacing also $h_{n-1}-h_n$ (see (\ref{h})) by $\mu-1$ the
series (\ref{ptilde}) can be summed and we obtain for (\ref{betacrit}):
\begin{equation*}
\frac{1}{\varepsilon} \int_0^\varepsilon \frac{1}{1+x-\mu} dx =
1 \ .
\end{equation*}
If $\varepsilon = 0$ this leads to the free gas
critical value $\mu=0$, but for $\varepsilon > 0$ we obtain
\begin{equation}\label{mu>0}
\mu = \frac{e^\varepsilon - 1 - \varepsilon}{e^\varepsilon - 1} > 0 \ .
\end{equation}
Similarly, the density equation (\ref{rhocrit}) now reads as
\begin{equation}\label{pho-eq-small-lambda}
\frac{1}{\varepsilon} \int_0^\varepsilon \frac{1}{e^{\beta
(1-\mu+x)} -1} dx = \rho \ .
\end{equation}
By (\ref{mu>0}) we can approximate for small $\varepsilon$ $\mu$ by
$\mu \approx \varepsilon/2$
and inserting it in (\ref{pho-eq-small-lambda}) we find
\begin{equation}\label{pho-eq-small-lambda-1}
 \frac{1}{\varepsilon} \int_0^\varepsilon \frac{1}{e^{\beta
(1 - \varepsilon/2 +x)} - 1} dx = \rho \ .
\end{equation}
By convexity of the function
$(e^{\beta(1+x)}-1)^{-1}$, we conclude for solution of the equation
(\ref{pho-eq-small-lambda-1})that
\begin{equation*}
\beta_c(\rho, \varepsilon) > \beta_c(\rho, 0) = \ln \left( 1 + \frac{1}{\rho}
\right) \ .
\end{equation*}
Notice that this argument also applies in the case of a discrete distribution, see Figure 5.

\section{Conclusion}

We conclude by few remarks concerning our results and open
problems. Summarizing the most striking observations about the
model considered in this paper, we have seen that at large values
of the on-site repulsion with a discrete distribution of the
random single-site particle potential, the disorder causes a
suppression of Bose-Einstein condensation at fractional values of
the density. On the other hand, the suppression of Bose-Einstein
condensation at integer values of the density observed in the
absence of disorder is lifted. For continuous distributions we
found that the critical temperature decreases with increasing
disorder for non-integer densities.

We have have concentrated here on the case of uniformly
distributed random external potential. Nonuniform distributions as
well as a random on-site interaction may also be of interest and
give rise to new phenomena. Of course, all our results concern the
infinite-range-hopping model. It would be of considerable interest
to extend our results to the short-range hopping model.

\vspace{1.0cm}

\textbf{Acknowledgements.} This paper is a result of a number mutual
visits: TCD would like to acknowledge an enjoyable visit to the
Centre de Physique Th\'eorique, Marseille, LAP and VAZ are thankful
to the  School of Theoretical Physics for hospitality and support
during their visits of Dublin Institute for Advanced Studies.



\newpage


\begin{thebibliography}{9999999}

\bibitem[MM]{Mat-Mat} T. Matsubara and H. Matsuda, A lattice model of liquid helium,
\textit{Progr. Theor. Phys.} \textbf{16}, 569--582 (1956).
\bibitem[U]{Ueltschi} D. Ueltschi, Geometric and probabilistic aspects of boson lattice models,
in \textit{Progr. Probab.} \textbf{51}: 363--391 (Birkh\"{a}user,
Boston, MA, 2002).
\bibitem[G-B]{GMEHB} G. Greiner, O. Mandel, T. Esslinger, T. W. H\"{a}nsch, and I. Bloch,
Quantum phase transition from a superfluid to a Mott insulator in
a gas of ultracold atoms, \textit{Nature} \textbf{415}: 39--44
(2002).
\bibitem[BD]{BD} J.-B. Bru  and T. C. Dorlas, Exact solution of the
infinite-range-hoping Bose-Hubbard model, {\it J.Stat.Phys.} {\bf
113}: 177--195 (2003).
\bibitem[A-Y]{ALSSY} M. Aizenman, E. H. Lieb, R. Seilinger, J. P. Solovej, and J. Yngvason,
Bose-Einstein quantum phase transition in an optical lattice
model, \textit{Phys. Rev.} A \textbf{70}: 023612-1--12 (2004).
\bibitem[KL1]{KL1} M. Kac and J. M. Luttinger, Bose-Einstein
condensation in the presence of impurities, \textit{J. Math.
Phys.} \textbf{14}: 1626--1628 (1973).
\bibitem[KL2]{KL2} M. Kac and J. M. Luttinger, Bose-Einstein
condensation in the presence of impurities II, \textit{J. Math.
Phys.} \textbf{15}: 183--186 (1974).
\bibitem[LS]{LS} J. M. Luttinger and H. K. Sy, Bose-Einstein
condensation in a One-Dimentional Model with Random Impurities,
\textit{Phys. Rev.}A \textbf{7}: 712--720 (1973).
\bibitem[L-Z]{LPZ} O. Lenoble, L. A. Pastur and V. A. Zagrebnov, Bose-Einstein
condensation in random potentials \textit{Comptes-Rendus de
l'Acad\'{e}mie des Sciences} (Physique) \textbf{5}: 129--142 (2004).
\bibitem[LZ]{LZ} O. Lenoble and V. A. Zagrebnov, Bose-Einstein condensation in the
Luttinger-Sy Model  (\textit{subm. to Markov Proc. \& Rel.
Fields}).
\bibitem[F-F]{FWGF} M. P. A. Fisher, P. B. Weichman, G. Grinstein and D. S. Fisher,
\textit{Phys. Rev.} B \textbf{40}: 456-470 (1989).
\bibitem[KT]{KT} M. Kobayashi and M. Tsubota, Bose-Einstein
condensation and superfluidity of a dilute Bose gas in a random
potential, \textit{Phys. Rev.}B \textbf{66}: 174516-1--7 (2002).
\bibitem[K-C]{KTC} W. Krauth, N. Trivedi and D. Ceperley, Superfluid-Insulator Transition in
Disordered Boson Systems, \textit{Phys. Rev. Lett.} \textbf{67}:
2303--2310 (1991).
\bibitem[K-S]{KLS} T. Kennedy, E. H. Lieb and B. S. Shastry, The X-Y model has long-range order
for all spins and all dimensions geater than one, \textit{Phys.
Rev. Lett.} \textbf{61}: 2582--2584 (1988).
\bibitem[AB]{AB} N. Angelescu and M. Bundaru, A Remark on the Condensation in the Hard-Core
Lattice Bose Gas, \textit{J. Stat. Phys.} \textbf{69}: 897--903
(1992).
\bibitem[B-T]{BBZKT} N. N. Bogolyubov (jr), J. Brankov, V. A. Zagrebnov, A. M. Kurbatov, and N. Tonchev,
Some classes of exactly soluble models of problems in Quantum
Statistical Mechanics: The method of approximating Hamiltonian, {\it
Russian Math. Surveys} {\bf 39}: 1--50 (1984).
\bibitem[PF]{PF} L. A. Pastur and A. L. Figotin, Theory of disordered spin systems,
{\it Theor. and Math. Phys.} {\bf 35}: 403--414 (1978).
\bibitem[PS]{PS} L. A. Pastur and M. V. Shcherbina, Infinite-range
limit for correlation functions of lattice systems, {\it Theor.
and Math. Phys.} {\bf 61}: 955--964 (1983).
\bibitem[RS2]{RS II} M. Reed and B. Simon, {\it Method of Modern Analysis II:
Fourier Analysis, Self-Adjointness} (Academic Press, New York,
1975).
\bibitem[ZB]{ZB} V. A. Zagrebnov and J.-B. Bru, The Bogoliubov model of weakly
imperfect Bose gas, {\it Phys.Rep.} {\bf 350}: 291--434 (2001).
\bibitem[D]{D} E. B. Davis, {\it One-Parameter Semigroups} (Academic Press, London,
1980).
\bibitem[NZ]{NZ} H. Neidhardt and V. A. Zagrebnov, Regularization and Convergence for
Singular Perturbations, {\it Commun. Math. Phys.} {\bf 149}:
573--586 (1992).
\bibitem[RS1]{RS I} M. Reed and B. Simon, {\it Method of Modern Analysis I:
Functional Analysis} (Academic Press, New York, 1972).
\bibitem[Z]{Z} V. A. Zagrebnov, {\it Topics in the Theory of Gibbs Semigroups},
Lecture Notes in Mathematical and Theoretical Physics, Vol.10
(Leuven University Press, Leuven 2003).
\bibitem[K]{K} T. Kato, \textit{Perturbation theory for linear operators}
(2nd ed, Springer Verlag, Berlin 1980).
\end{thebibliography}
\end{document}